\documentclass[a4paper,onecolumn,12pt,journal,draftclsnofoot]{IEEEtran}

\usepackage[latin2]{inputenc}
\usepackage{graphicx}
\usepackage{amstext}
\usepackage[]{ntheorem}
\usepackage{fancyhdr}
\usepackage{amsfonts,dsfont}
\usepackage{amsmath,amssymb}
\usepackage{t1enc}
\usepackage{algorithm}
\usepackage{algpseudocode}
\usepackage{multirow}
\usepackage{subcaption}

\begin{document}

%--------------------------------------------------------------------------------

\title{Preallocation-based Combinatorial Auction for Efficient Fair Channel Assignments in Multi-Connectivity Networks}
% A preallocation based combinatorial auction channel allocation algorithm in multi-connectivity networks
%\title{Comparison of algorithmic game-theory based channel allocation mechanisms in multi-connectivity networks}

\author{D\'{a}vid Csercsik$^1$ and Eduard Jorswieck$^2$}
\date{}

\maketitle

\vspace{-1cm}
\begin{abstract}
\footnotetext[1]{Centre for Economic and Regional Studies of the Hungarian Academy of Sciences, T\'oth K\'alm\'an u. 4., H-1097 Budapest, Tel.: +36-1 309 26 52, Fax: +36-1 319 31 36 and P\'{a}zm\'{a}ny P\'{e}ter Catholic University, Faculty of Information Technology, P.O. Box 278, H-1444 Budapest, Email: \texttt{\small csercsik.david@.krtk.hu}

This work has been supported by the Hungarian Academy of Sciences under its Momentum Programme LP2021-2 and by the Fund and by K 131 545 
of the Hungarian National Research, Development and Innovation Office. 
%This work has been supported by the Fund and by K 131 545 of the Hungarian National Research, Development and Innovation Office.
D\'{a}vid Csercsik is a grantee of the Bolyai scholarship program of the Hungarian Academy of Sciences.}

\footnotetext[2]{TU Braunschweig, Institute for Communications Technology, Schleinitzstr. 22, 38106 Brunswick, Germany,
Email: \texttt{\small jorswieck@ifn.ing.tu-bs.de}

The work of E. Jorswieck is partly funded by the German Research Foundation (DFG) under grant JO 801/24-1 and by the Federal Ministry of Education and Research (BMBF) within the 6G Research and Innovation Center (6G-RIC) under support code 16KISK031.}

We consider a general multi-connectivity framework, intended for ultra-reliable low-latency communications (URLLC) services, and propose a novel, preallocation-based combinatorial auction approach for the efficient allocation of channels. We compare the performance of the proposed method with several other state-of-the-art and alternative channel-allocation algorithms. The two proposed performance metrics are the capacity-based and the utility-based context. In the first case, every unit of additional capacity is regarded as beneficial for any tenant, independent of the already allocated quantity, and the main measure is the total throughput of the system. In the second case, we assume a minimal and maximal required capacity value for each tenant, and consider the implied utility values accordingly. 
In addition to the total system performance, we also analyze fairness and computational requirements in both contexts.
We conclude that at the cost of higher but still plausible computational time, the fairness-enhanced version of the proposed preallocation based combinatorial auction algorithm outperforms every other considered method when one considers total system performance and fairness simultaneously, and performs especially well in the utility context. Therefore, the proposed  algorithm may be regarded as candidate scheme for URLLC channel allocation problems, where minimal and maximal capacity requirements have to be considered. 
\end{abstract}

%\textbf{Keywords}: Resource allocation, multi-connectivity, channel-assignment, URLLC, combinatorial auction

% I think we need to enter them in the submission process - do not have to be here

%=========================================================
%\newpage

\section{Introduction}

Ultra-reliable low-latency communications (URLLC) is defined as one of the fundamental requirements in the case of emerging fifth generation (5G) and beyond mobile wireless communications systems \cite{chen2018ultra}. Enabling URLLC is essential in mission-critical applications, like wireless factory automation, coordination of vehicles and real-time remote control applications in the internet of things \cite{Hossler2018Mission}. 

The current 5G New Radio (NR) is the first wireless standard designed to natively support multi-service communications. In addition to serving conventional HTC through the enhanced mobile broadband (eMBB) service class, 5G NR supports MTC through the newly introduced ultra-reliable low-latency communications (URLLC) and massive MTC (mMTC) service classes. Among the three 5G NR service classes, URLLC is arguably the most challenging.
In 5G NR, this challenge is mainly addressed by a new 5G numerology, i.e., differentiated parameter choices for the orthogonal frequency division multiplex (OFDM) access \cite{Segura2021}. As outlined in the conclusions of \cite{Segura2021}, the reliability needs further improvements which cannot be realized by the new numerology alone. 

As the redundancy of communication channels improves the reliability of the communication architecture, using multiple paths at once for communicating objects (\emph{tenants} in the following) is a potential tool to achieve URLLC requirements \cite{suer2019multi}.
However, as the resources are finite, the allocation of channels to individual tenants poses a challenge in such multi-connective, multi-user environments. Furthermore, the computational complexity of the underlying assignment problem is growing with the number of tenants as well as the number of communication paths \cite{Manap2020}. Therefore, deep learning methods are proposed to solve these challenging resource allocation problems \cite{She2021}. 

In addition to machine learning, game-theory based algorithms are discussed for resource allocation in 6G \cite{Elsayed2019}. Using game theoretic approaches for resource allocation problems in wireless systems has an extensive literature (for a review see \cite{CHARILAS20103421}).
Most of these approaches are based on the framework and tools of non-cooperative game theory \cite{fudenberg1991game}, using its relatively widely known concepts
as the Nash-equilibrium \cite{nash1951non} or the subgame-perfect equilibrium \cite{fudenberg2009subgame}.  One may find also examples of cooperative \cite{saad2009coalitional,csercsik2017cooperation} and division-theory based game-theoretic applications as well \cite{csercsik2019cake}.

Algorithmic game theory \cite{nisan2009algorithmic,roughgarden2010algorithmic} on the one hand studies the concepts of the previously mentioned branches of game theory in an algorithmic context, and on the other hand it analyzes further algorithm-focussed social choice and economic problems like the stable marriage problem \cite{gale1962college}, the house-allocation problem \cite{ALCALDEUNZU20111} or the problems related to the design of auctions and mechanisms \cite{roughgarden2019approximately}.

The channel allocation problem may be regarded as a special case of spectrum sharing, where we assume that prior defined, indivisible resources (channels) have to be allocated to tenants. The assumption of indivisibility may be relaxed if we consider time-sharing, but we do not focus on such cases in the current paper. In the multi-connective case, the number of channels assigned to a particular tenant may also vary. Methods related to the field of algorithmic game theory, like the Gale-Shapley (GS) or delayed acceptance algorithm have been already proposed to similar problems \cite{jorswieck2013matching,Hossler2019matching}.

% intermittent communications 

%@ARTICLE{8306104,
%  author={Habiba, Ummy and Hossain, Ekram},
%  journal={IEEE Communications Surveys   Tutorials}, 
%  title={Auction Mechanisms for Virtualization in 5G Cellular %Networks: Basics, Trends, and Open Challenges}, 
%  year={2018},
%  volume={20},
%  number={3},
%  pages={2264-2293},
%  doi={10.1109/COMST.2018.2811395}}

%@INPROCEEDINGS{9155532,  author={Qin, Qiaofeng and Choi, Nakjung and Rahman, Muntasir Raihan and Thottan, Marina and Tassiulas, Leandros},  booktitle={IEEE INFOCOM 2020 - IEEE Conference on Computer Communications},   title={Network Slicing in Heterogeneous Software-defined RANs},   year={2020},  volume={},  number={},  pages={2371-2380},  doi={10.1109/INFOCOM41043.2020.9155532}}

%\subsection{Contributions and structure of the paper}

The major contributions of our work are summarized as:
\begin{itemize}
    \item We propose two novel channel allocation methods for multi-connectivity networks, based on the principles of the combinatorial auction problem \cite{de2003combinatorial}, one with enhanced fairness properties. 
    \item In order to achieve computational feasibility, we apply a preallocation step for the combinatorial auction. 
    \item We compare various algorithmic game theory-based channel assignment algorithms with respect to capacity and utility performance. 
    Thereby, we develop a descriptive framework, in which the different channel allocation methods are evaluated and consider the special case of URLLC multi-connectivity. 
\end{itemize}
%In this work we aim to compare various algorithmic game theory-based channel assignment algorithms in the multi-connective context, and propose a novel channel allocation method, based on the principles of the combinatorial auction problem \cite{de2003combinatorial}. In order to do this, we first define a descriptive framework, in which the different channel allocation methods may be evaluated. As a particular case of the defined general framework, we consider the URLLC multi-connectivity model in the study. 

The structure of the paper is as follows:
Section \ref{section_model} of the paper describes the system model and the used evaluation metrics as well as the problem statement, Section \ref{sec_CA_algorithms} introduces the analyzed channel assignment algorithms, and introduces the proposed solution method, Section \ref{sec_results} summarizes the numerical simulation results, Section \ref{sec_discussion} provides additional discussion of the results and Section \ref{sec_conclusions} concludes.

%%%

\section{Connectivity model and problem statement}
\label{section_model}

\subsection{General setting of the multi-connectivity channel assignment problem}
\label{subsec_connectivity_model}

The key assumptions of our modelling framework are the following.
In the modelled area a finite number of base stations (BSs) is present, each providing a finite number of available channels. We denote the $j$-th channel of BS $i$ by $ch_{i,j}$, and the set of all channels by $CH$. \footnote{The cardinality of $CH$ is denoted by $|CH|$, and the set of all subsets of $CH$ is denoted by $2^{CH}$.} If we consider a general subset of channels ($S$), where the individual channels not necessarily belong to the same BS, we index the channels by $m$.
  $n_{ch}^i$ denotes the number of channels offered by BS $i$, while $n_{ch}$ denotes the total number of channels (i.e. $\sum_i n_{ch}^i = n_{ch}$).

A finite number of tenants is present in the modelled area. Tenant $k$ is denoted by $T_k$. $n_T$ denotes the number of tenants. Channels may be assigned to tenants. In the multi-connective framework, each channel may be assigned to maximum one tenant, but one tenant may hold multiple channels. This refers to many-to-one matching market.\footnote{We restrict the model to exclusive assignments of channels to tenants. Note, that it might be also possible to support time-sharing where a channel is partly used by multiple tenants.} Let us introduce the binary assignment matrix $A \in \mathcal{B}^{n_T\times n_{ch}}$. $A(k,m)=1$ iff $ch_m$ is assigned to $T_k$, else $A(k,m)=0$. Let us note, that the columns of $A$ may be grouped according to the relevant BS of the actual channel. The row $k$ of $A$ determines the set of channels assigned to $T_k$, denoted by $S_k$.

The set of assigned channels explicitly determines the (bit)rate of any tenant, and this rate does not depends on the channel assignment of other tenants. This assumption is related to the phenomenon of co-channel interference. In the context of the current paper we do not assume that co-channel interference is neglected, but we assume that it is bounded and its maximal potential effect is considered in a term, which is independent of the assignment setup. This worst case interference model is applied in many recent works, e.g. \cite{Shakir2014}. 
  % @EJ: I added the latter two sentences here, please verify.
  
The connectivity function\footnote{Not to be confused with the connectivity function used in discrete mathematics \cite{carlson2006gene,seymour1988connectivity}.} of tenant $k$ is denoted by $\rho_k:~2^{CH} \rightarrow \mathcal{R} $. $\rho_k(S_k)$ describes the resulting rate of tenant $k$ if the subset $S_k$ of all channels is assigned to it. The assumption that the value of $\rho_k(S_k)$ depends only on $S_k$ and not on other $S_p$-s, where $p \neq m$. In other words, interference phenomena caused by the allocation of other available channels is neglected. Considering the beamforming features allowed by novel massive MIMO techniques, such as zero-forcing beamforming,  \cite{guo2014waveform,bjornson2019massive}, this assumption may be regarded as plausible in the current technological environment. 
  % move or add arguments about interference 

The concept of the connectivity function allows us to use a modular structure. As we define each channel assignment algorithm based on $\rho$, it is possible to modify the underlying connectivity model (including e.g. the signal propagation model), and use the same algorithms without modification.
In the following, we discuss the details, based on which the particular connectivity function used in this study may be derived.

\subsection{SIR-based connectivity function}
\label{subsec_used_conn_fncn}

The connectivity function used in this paper is derived following the principles of the URLLC framework described in \cite{Hossler2019matching}.
The model is based on the following assumptions.

\subsubsection{Single connectivity}
Let us first discuss the special case, when the tenant in question is connected only to a single channel of a particular BS.
We denote the local mean signal-to-interference ratio (SIR) in dB of tenant $k$ operating using the $j$-th channel of BS $i$ with $\bar{\gamma}_{(k,i,j)}$ ($S_k=ch_{i,j}$). This quantity may be calculated as $\bar{\gamma}_{(k,i,j)}=P^{R}_{(k,i,j)}-P^I_{k,j}$, 
where $P^I_{k,j}$ corresponds to the co-channel interference of channel $j$ observed at tenant $k$ and $P^{R}_{(k,i,j)}$ denotes the received power, which may derived as
 \begin{equation}
P^{R}_{(k,i,j)}=P^T_{i,j} - PL(d_{k,i}).
\label{P_R}
\end{equation}
$P^T_{i,j}$ in (\ref{P_R}) denotes the transmission power of BS $i$ on channel $j$, while $PL(d_{k,i})$ is the path loss in dB, assuming the distance $d_{k,i}$ between the tenant $k$ and BS $i$. Here, it is assumed that one BS serves at most one user on channel $j$. The path loss $PL(d_{k,i})$ is calculated as follows
$PL(d_{k,i})=PL(d_0) + 10\delta \log_{10}\left( \frac{d_{k,i}}{d_0} \right)$, where $\delta$ denotes the path loss exponent, and $PL(d_0)$ stands for the reference path loss, valid in the case of the reference distance $d_0$.

In addition, we can compute that the instantaneous channel capacity ($C$) achieved by tenant $k$ on channel $j$ with BS $i$, as
\begin{equation}\label{ch_capacity}
  C_{k,j,i} =B \cdot \log_2 (1+\gamma_{k,i,j}),
\end{equation}
where $B$ denotes the bandwidth, and $\gamma_{k,i,j}$ is the random instantaneous SIR. If we consider a slow fading channel, an outage event occurs, when the required data rate $R_k$ of tenant $k$ exceeds the capacity 
$P^{out}_{k,j,i}=Pr[C_{k,j,i}<R_k]=Pr[\gamma_{k,j,i}<\gamma^{th}_{k}].$ 

% CsD: According our telco, if I understood right, e currently do not apply this, but this corresponds to an other potential connectivity model
%(\ref{outage}) describes that outage occurs if $\gamma$ is less than a threshold $\gamma_{th} = 2^{\frac{C}{B}}-1$.
% Yes, your interpretation of the outage probability and gamma threshold is correct.
A typical measure applied by operators is the so called outage capacity, i.e. the highest possible transmission rate, which keeps the outage probability below a certain value $\varepsilon$, i.e., for tenant $k$ scheduled on channel $j$ at BS $i$, it reads $\rho_{k,j,i} = C^{\varepsilon}_{k,j,i} = \max \{ R_k : P^{out}_{k,j,i} < \varepsilon \}.$

According to \cite{yao1990outage}, assuming Rician/Rayleigh fading environment and a single interferer, the outage probability of channel $m$ may be calculated as
\begin{equation}
P^{out}_{k,j,i}=\frac{\gamma^{th}_k}{\gamma^{th}_k + \bar{\gamma}_{(k,j,i)}}~ \text{exp} \left( -\frac{K_{k,i} \bar{\gamma}_{(k,j,i)}}{\gamma^{th}_k +\bar{\gamma}_{(k,j,i)}} \right)
\label{P_out_single},
\end{equation}
where the Rician factor $K_{k,i}$ characterizes the ratio between the power in the dominant path of the desired signal and the power in the scattered paths of the interferer, regarding the tenant-BS pair $(k,i)$.

\subsubsection{Multi-connectivity}

Let us extend this to the case where tenant $k$ is assigned a set of channels, denoted by $S_k$, we obtain the resulting $\varepsilon$-outage capacity as  connectivity function 
\begin{equation}\label{outage_capacity}
\rho_k(S_k):=C^\varepsilon_{S_k} =\text{max}~ \{R_k:P^{out}_{S_k}<\varepsilon \}.
\end{equation}
In order to compute (\ref{outage_capacity}), we need a model for the outage probability of assigned a set of channels to one tenant. The resulting outage probability depends on the combining scheme and the joint distribution of the underlying fading processes of the channels. Potential combining schemes are maximum ratio combining, equal gain combining or selection combining. While our proposed framework is able to support all of them, we consider here selection combing. The proposed framework can handle general joint distributions, here we assume statistically independent Rician/Rayleigh fading channels. We chose selection combining and independent Ricean/Rayleigh fading, in order to focus on the diversity effect of multi-connectivity. 

The outage probability considering a subset $S_k$ of channels (potentially belonging to different BSs), is described in (\ref{P_out_multic}) under the assumption that they are statistically independent. 
\begin{eqnarray} \nonumber 
P^{out}_{k} & =& \prod_{(j,i) \in {S_k}} P^{out}_{k,j,i}   = \prod_{(j,i) \in {S_k}}\frac{\gamma^{th}_k}{\gamma^{th}_k + \bar{\gamma}_{k,j,i}}~ \text{exp} \left( -\frac{K_{k,i} \bar{\gamma}_{k,j,i}}{\gamma^{th}_k + \bar{\gamma}_{k,j,i}} \right).
\label{P_out_multic}
\end{eqnarray}

Following these principles, $\rho_k(S_k)$ is determined as follows. We consider the  (\ref{P_out_multic}) assuming $L=|S_k|$, and (considering the respective $\bar{\gamma}_m$ parameters) numerically determine the maximal value of $\gamma_{th}$ for which the inequality described in (\ref{outage_capacity}) for fixed $\epsilon$ holds. Formally, this corresponds to finding the inverse function of the outage probability which can be written as an explicit function of the rate $R_k$, i.e., solve $P^{out}_k(R)=\varepsilon$ for $R_k$: $\rho_k(S_k) = \left( P^{out}_k \right)^{-1}(\varepsilon)$. 
The invariant parameters of the used model are described in \ref{subsec_simulation_setup}. As we consider $\varepsilon=10^{-9}$ in all simulations, the upper index $\varepsilon$ is omitted in the notation in the following.

On the other hand, the following model parameters are varied during the simulation scenarios. (1) number and geometric position of tenants, (2) number, geometric position, transmit power of base stations ($=P^T_i$ for BS $i$) and (3) number of channels available per base station.
%  \item Transmission parameters

%In the next section we describe the various channel assignment algorithms studied in the evaluation process.

\section{The channel assignment problem and assignment algorithms}
\label{sec_CA_algorithms}

In this section we discuss the principles of the allocation process, and enumerate the channel allocation algorithms which are evaluated in this study.

\subsection{Capacity versus utility-based assignment}
\label{cap_vs_ut}

As described in Subsection \ref{subsec_used_conn_fncn}, in the basic setup, $\rho_k(S_k)$ equals to the resulting channel capacity value $C$, assuming that the set $S_k$ of channels is assigned to tenant $k$. Considering only the capacity of a given allocation as a guideline for the assignment, or considering the total resulting capacity of the system as a sole measure of performance neglects some aspects of URLLC wireless communications.

The first aspect is the potential minimal capacity requirement of tenants, the so-called Quality of Service (QoS) requirements. In the case of practical applications like real time control or factory automation, the assumption of a minimal rate required for standard operation is straightforward. This aspect is connected to the fairness of the channel allocation, namely that while we maximize the system performance in the terms of total capacity, some tenants may remain without assigned channels, or with low resulting capacity. This aspect is already discussed in the literature, see e.g. the papers \cite{Hossler2019matching,simsek2019multiconnectivity} related to URLLC modelling frameworks.

The second aspect to be considered is the dual of the first. It is plausible to assume that after a certain level, additional units of capacity do not bring benefits to any tenant. The usual traffic in the case of e.g. factory automation applications, does not exceeds a certain limit, thus too high capacities will be potentially left unused.

To address these two considerations, we use the concept of the \emph{utility function}. The utility function of tenant $k$, denoted by 
$U_k(C_k): \mathcal{R}^+ \rightarrow [0,1]$ assigns a normalized utility value, depending on the actual capacity value of the tenant $k$, defined by the current channel allocation ($C_k=\rho_k(S_k)$). It is furthermore plausible to assume that such utility functions are monotone increasing.

The paper \cite{SHI20082257} introduces various utility functions for the description of network users based on the nature of the user. The 'TCP interactive user' profile defined in this paper matches our above assumptions in the sense that the resulting utility is considered as 0 below a minimal capacity value ($C_k^{min}$) of the current user, and shows a monotone concave increase between the minimal and the maximal ($C_k^{max}$) capacity value. Above the maximal capacity value, the function is saturated and additional capacity does not bring additional benefits (i.e. increment in the utility function). 

Based on the 'TCP interactive user' profile of \cite{SHI20082257}, we use the utility function described by eq. (\ref{utility_fncn}) in the current paper.

\begin{equation}\label{utility_fncn}
U_k(C_k)=
\begin{cases}
\frac{\log(C_k/C_k^{min})}{\log(C_k^{max}/C_k^{min})}\frac{\textrm{sgn}(C_k-C_k^{min})+1}{2} &\text{if}~  C_k \leq C_k^{max}\\
1 &\text{if}~ C_k>C_k^{max}\\
\end{cases}
\end{equation}

The function described in e.q. (\ref{utility_fncn}) holds two parameters, namely $C_k^{min}$ and $C_k^{max}$ for each tenant. An example case is depicted in Fig.\ref{utility_fncn_example} of Appendix A.

During the simulation of the algorithms we will analyze their performance in capacity or utility context. In the former case we will assume that the decisions of the participants and the outcome of algorithms are determined according to the respective values of the connectivity function, while in the latter case the implied utility values will serve as basis for the preferences, evaluations and decisions during the assignment process.

As it has been discussed earlier, in the utility context, the implied utility values serve as basis for preferences, evaluations and decisions. However, most of the allocation methods is based on preference lists of tenants over single channels. In the proposed simulation setup discussed later, the allocation of a single channel usually does not provide enough capacity for the tenant to reach the $C^{min}$ value, thus if we evaluate the utility implied simply by the single connectivity capacity values, comparison, and thus setting up preference list may be problematic (as lots of the values are 0).
To resolve this issue we use the following approach. If preferences over single channels are to set up in the utility context, and a tenant receives its first channel, we evaluate the utility improvement implied by the capacity ensured by the respective channel, using $C^{min}_k$ as reference. In other words, if channel $m$ ensures $C_{m,k}$ capacity for tenant $k$ in the case of single-connectivity (if no other channel is used), the value serving as basis for comparison is $\rho_k(C^{min}_k+C_{m,k})$. If preferences over single channels are evaluated, but the tenant already has one or more channels allocated (such scenarios arise in the case of WS, ORR, MRM and MRGS), the preferences are set up based on the utility improvement implied by the potential additional channel.

\subsection{The channel assignment problem}
According to the above considerations, the general channel assignment problem may be described as an optimization problem formulated in eq. (\ref{eq_assignment_opt}), where the aim is to maximize the overall capacity or total utility of the system (depending on the actual context).
\begin{eqnarray}
    & \max_{A} \sum_{k=1}^{n_T} \rho_k(S_k)
    %\quad (\sum_{k=1}^K \rho_k(S_k))
    \text{ or }
    \max_{A} \sum_{k=1}^{n_T} U_k(\rho_k(S_k))
    \nonumber \\
    & \text{s.t. } \sum_k A(k,m) \leq 1 ~~~ \forall m ~(1 \leq m \leq n_{ch})
    \label{eq_assignment_opt}
\end{eqnarray}
In this formulation, the $k$-th row of $A$ may be considered as the membership function of the set $S_k$: $A(k,m)==1$ iff $ch_m$ is assigned to $T_k$, i.e. iff 
$ch_m \in S_k$.  The formulation (\ref{eq_assignment_opt}) defines an integer programming problem (or a combinatorial optimization problem, since the space of the possible solutions is finite), where the objective function is nonlinear.

In the case of realistic applications where the number of tenants and channels is medium or high (e.g. above 5 tenants and 15 channels), the combinatorial explosion of the search space practically  makes the problem (\ref{eq_assignment_opt}) computationally impossible to solve analytically.

The general approach to overcome this obstacle is to use heuristic based algorithms for the channel assignment procedure, like the weakest selects \cite{rhee2000increase} or the opportunistic round robin \cite{kulkarni2003opportunistic,hassel2006spectral} method, or to adapt matching algorithms originally designed for different context (like the famous Gale-Shapley method \cite{gale1962college}) for the problem.

These heuristics and adaptations may be evaluated against each other according to the objective function of the formulation (\ref{eq_assignment_opt}), which is related to the overall performance of the system, and to different other aspects like fairness, starvation rate, and similar measures.

Note that time sharing between different assignment strategies could relax the programming problem in (\ref{eq_assignment_opt}). Since we consider non-elastic low latency traffic, we do not consider time sharing as an option in this work. 
% Remark that in the current paper we don't consider time sharing, but if we'd do it would relax the optimization problem 

\subsection{Randomized assignment methods}

% more formal description? 
These methods do not consider the values of the connectivity function in the assignment process.

\subsubsection{Random assignment (\textbf{R})}

This assignment method is used as a dummy reference case. The process is very simple. We define a maximal possible channel number per tenant, denoted by $\bar{n}_{ch}$.
Following this we consider the available channels one by one. For each channel we pick a random tenant (according to uniform distribution), and if the tenant doesn't already has
$\bar{n}_{ch}$ channels assigned to it we assign the actual channel to the tenant. If the tenant is already full, we choose an other tenant.

\begin{algorithm}
\caption{Random assignment (R)}\label{alg:R}
\begin{algorithmic}
\State $A \gets 0^{n_T \times n_{ch}}$   
\For{$i = 1:n_{ch}$}
\State $k_R \gets RI(1,n_T)$ \Comment{A random integer between 1 and $n_T$ is generated uniformly}
\If{$\sum_m A(k,m)< \bar{n}_{ch}$}
    \State $A(k_R,i) \gets 1$
\EndIf
\EndFor
\end{algorithmic}
\end{algorithm}

\subsubsection{Distance based semi-random assignment (\textbf{SR1})}

% add a reference   / OK 
This method is a slightly improved version of the \textbf{R} method. The only difference is that to avoid extremely inefficient assignment setups, channels are assigned to closer tenants with higher probability\footnote{Note that the concept of using channel quality to influence the probability of access was proposed for channel-aware medium access control protocols before \cite{Jin2012}.}. More precisely, the probability of assignment is proportional to $1/d$. In other words, if we would like to allocate a channel to one of two tenants, from which one is 20m away from the BS offering the channel and the other is 40m away, the first tenant is twice more likely to receive the channel in the case of \textbf{SR1}. This modification avoids very unlucky matchings with high probability, but still does not consider any preferences or tenant capacity requirements explicitly.

\subsubsection{Single-connectivity value based semi-random assignment (\textbf{SR2})}

The method is a straightforward modification of the SR1, but in this case not the respective Tenant-BS distances are used as reference for the assignment, but the respective capacity values, resulting from the single connectivity scenarios, if the Tenant is connected only via the channel in question. In contrast to the SR1, this method also implicitly considers the effect of the potentially various transmission power values ($P^T$) of base stations, and the actual Rician factors for the BS-tenant pairs ($K_{k,i}$).

\subsection{Selection based algorithms}
These algorithms allow tenants to choose from the available channels, according to the improvement implied by the increase in the value of the connectivity function or the implied utility (depending on context) for each alternative. The difference between these algorithms is the order, according to which we let the tenants choose from the still available channels.

\paragraph*{Random Tie-breaking}
In the case of selection based algorithms, tenants usually set up preferences over the available channels (and in some cases vice versa as well).

As we assume BSs potentially with multiple identical channels, the the problem of tie-breaking arises. In other words, if a BS offers multiple channels, with uniform parameters, the tenants don't have any reason to prefer any of them more than any other. In such cases, we always apply random tie-breaking, in other words in general, if two ore more channels/tenants are preferred at the same level according to quantitative indicators by tenants/channels (like distance of the corresponding BS/tenant, or the capacity/utility improvement ensured by the channel), the preference ordering of the respective items will be set up randomly.

\subsubsection{Weakest selects algorithm (\textbf{WS})}

This method, proposed in \cite{rhee2000increase}, already takes into account $\rho_k(S_k)$ or $U_k(\rho_k(S_k))$ in the allocation process.
The method implements a simple iterative process.

If the capacity context applies, in each step the tenant with the lowest capacity value chooses the channel, which ensures the most capacity improvement for it. 
In the spirit of random tie-breaking, if multiple such tenants are present or multiple channels are considered as best options, the algorithm determines the order of choice randomly.

If the utility context applies, the tenant with the highest utility deficit ($\Delta U_k(C_k)=1-U_K(C_k)$) chooses the channel, which ensures the most utility improvement for it. 

\begin{algorithm}
\caption{Weakest selects (WS) (in capacity context)}\label{alg:WS}
\begin{algorithmic}
\State $A \gets 0^{n_T \times n_{ch}}$   
\State $f_{acp} \gets 1$   \Comment{Flag showing that at least one available channel is present}
\While{$f_{acp} \neq 0$}
\State {Determine the sets $S_k$ based on $A$ for each $k$}
\State {Determine the $\rho_k(S_k)$ values for each $k$}
\State {Find the tenant(s) with minimal $\rho_k$ value. If multiple such tenants are present, choose one at random ($w$)}
\State {Find the most preferred channel of $T_{w}$. A channel denoted by $ch_m$ is regarded as most preferred if in the case of $S_k'=S_k \bigcup ch_m$, and $S_k''=S_k \bigcup ch_l$,~ $\rho_k(S_k'') \leq \rho_k(S_k') \text{ or } U_k(\rho_k(S_k'')) \leq U_k(\rho_k(S_k'))~~~\forall l \neq m$. If multiple such channels are present choose one at random}
\State $A(w,m)\gets 1$
\EndWhile
\end{algorithmic}
\end{algorithm}

\subsubsection{Opportunistic round robin algorithm (\textbf{ORR})}

This method is very similar to the \textbf{WS}, the only difference is that instead of looking always for the 'weakest' tenant, the tenants choose in a random order in each round. If there are still available channels after the actual round is finished, an additional round follows. The principle of the method has been proposed in the case of time-slot scheduling in \cite{kulkarni2003opportunistic,hassel2006spectral}.

\subsection{Methods based on the Gale-Shapley algorithm}

The Gale-Shapley (GS), or 'delayed acceptance' algorithm \cite{gale1962college} has been originally proposed for the stable marriage problem, and assumes that the elements of two disjoint sets (man and woman) have preferences defined over the elements of the other set (each man has his own preference ordering over women, and each woman has her own preference ordering over men). The algorithm pairs men with women via the following simple iterative method (we consider the 'men propose' version here).
\begin{enumerate}
  \item Every man proposes to the most preferred woman, who has not yet rejected him.
  \item Considering the incoming offers, every woman chooses the most preferred men, puts him on hold, and rejects the rest.
\end{enumerate}

If the number of men and women are equal, and each participant has a full preference list, the algorithm stops, when each women has exactly one offer.
This will result in a \emph{stable} matching, which means that considering the resulting pairs, no man and woman may be found, who are not paired together, but prefer each other more compared to their actual partner (i.e. no blocking pair is present).

A straightforward generalization of the problem and the algorithm is the many-to-one case, which corresponds to e.g. college admission problems. The role of the 'men' set here is played by the students, who apply for admission (in this case each student applies to one collage), and the set of collages play the role of the 'women' set, but with quotas. The difference to the basic algorithm is, that each collage $i$ puts the $q_i$ most preferred student on hold en each step, and rejects the rest (where $q_i$ is the quota of collage $i$).
The pseudo code of the many-to-one case may be found in \cite{Hossler2019matching}. 
The problem may be generalized even further, the many-to-may case is described in \cite{baiou2000many}.

Let us emphasize here, that while the GS method has been adapted to various resource allocation problems (see the references later), the principle of the algorithm is based on the concept of the preference lists. In other words, these approaches assume that individual indivisible resources are compared with each other, and bundles of these resources are not evaluated. In other words, we may know for example that the preference of a tenant over 4 channels, which are allocated simultaneously is $ch_1 \succ ch_2 \succ ch_3 \succ ch_4$, but we do not know if the tenant values the bundle $\{ch_1,~ch_4\}$ or $\{ch_2,~ch_3 \}$ more. In multi-connective environments, where the value of the connectivity function $\rho_k$ is determined based on the actual subset of assigned channels ($S_k$), this aspect arises as a clear limitation of the method. An other limitation of the method is that the preference lists are not quantitative. It not known e.g. how much $ch_1$ is preferred to $ch_2$ compared to the relation between $ch_2$ and $ch_3$.

Stability of the resulting matching is a critical and central aspect in the original problem, and the most significant virtue of the GS is that it results in a stable matching. While in general a stable matching is not unique (see e.g. \cite{palmer2020most}), there are special cases, when uniqueness holds \cite{clark2006uniqueness}. For example, it has been shown that there is a unique stable matching if the sets of men and women can each be ordered so that any man and woman with the same rank prefer each other above any other partner with a lower rank \cite{eeckhout2000uniqueness}. Capacity values resulting from single connectivity may serve as basis for such symmetric ranking, which already guarantees the uniqueness of stable matching.

Regarding wireless applications, stability of the matching may be important in decentralized allocation methods \cite{Gu2015}. In the current work, we assume that the notion of stability is not an explicit requirement, but it leads to fair outcomes of the assignment process \cite{Hoesler20}. Therefore, we apply the GS algorithm as a candidate approach. 

% (EJ) % decentralized assignments require stability 
% include papers by Amir Leshem - no, I added an overview paper.
% argument that we do not consider stability, but "apply" the algorithm to find an assignments.

Examples of GS (or delayed acceptance)-based applications in wireless resource sharing may be found in
\cite{Jorswieck2011,jorswieck2013matching,tahir2015coalition,tomasin2015gale,chang2016gale,butt2018spectrum}, while GS approaches for the the channel allocation problem in URLLC context are described in \cite{Hossler2019matching,simsek2019multiconnectivity,lee2020many}. In this paper, we consider 3 versions of the GS for the channel allocation problem.
In all cases, the preferences of the channels over tenants and tenants over channels are set up according to the capacity or utility values which are provided by single channels (single-connectivity values).

\subsubsection{Basic Gale-Shapley (\textbf{GS})}

Here we consider the many-to-one case. Channels are corresponding to men, and tenants correspond to woman, who have a quota greater than 1.

% single versus multiple stable matchings 

\subsubsection{Minimum-rate matching (\textbf{MRM})}

This version has been proposed in \cite{Hossler2019matching}. The principle of the method is the following. A minimum rate or utility value is defined for each tenant. Until the capacity/utility of any tenant is below this minimum rate, the WS algorithm is applied, with the modification that in each round the tenant with the highest capacity/utility deficit chooses its most preferred channel.
If the minimum values are achieved in the case of every tenant, the remaining channels are allocated via the GS.

\subsubsection{Multi-Round Gale-Shapley (\textbf{MRGS})}

It is also possible to apply the GS in an iterative manner. In this version, multiple rounds are performed, and each BS offers only one channel in each round. If a BS runs out of channels, it does not participate in the following rounds. This method, similar to the MRM, may be regarded also as a fairness enhanced-version of the GS, since as long as the total number of offered channels (equal to $n_{BS}$) is at least equal to $n_T$, every tenant receives exactly one channel in each round (not allocated channels are offered in the next round). 

In addition, tenants update their preferences over channels after each round in this method, according to the additional benefit the individual additional channels would bring. In the utility context this may e.g. imply that tenants who are already close to $C^{max}_k$ may be equally satisfied with any additional channel.
A similar approach is applied in \cite{quintiliireaching}.

\subsection{Method of the top trading cycles (\textbf{TTC})}

The method of the top trading cycles (TTC) has been proposed originally for the housing market \cite{ALCALDEUNZU20111}. In this case we assume $n$ owners, each owing a 'starting' house, and having a preference list over houses.
The algorithm evolves as follows.  In the first step, each owner points to the house most preferred by him/her. This way a directed graph is defined, where the out-degree of every node is 1.  Such graphs always hold at least one circle (loop edges are possible, and regarded as circles here). In the second step, we exchange the houses along the circles, and following this we remove the owners and houses included in the exchange transactions.
We repeat this process, until each house is allocated.

The method may be applied to the channel allocation problem in a way similar to the MRGS algorithm. In each round we consider a set of channels to be allocated, with cardinality equal to the number of tenants (preferably by different BSs). Initially, we distribute these channels among tenants by random, and we perform the TTC to get the resulting allocation of the actual round. In the next round we consider a set from the remaining channels and so on, until each channel is allocated.
 The preferences of tenants over channels are always set up according to the value of capacity or utility improvement which are implied by the potential channels (considering the channels already allocated).

According to our best knowledge, the TTC algorithm has not been applied to
wireless resource allocation problems. However, as we will see later, the principle of the algorithm (exchanges are performed only if they are beneficial for all the involved parties) limits its performance in the application currently considered.

\subsection{Methods based on combinatorial auction}

As described in \cite{de2003combinatorial}, the principle of combinatorial auction (CA) is that every participant may place bids for multiple (or in the original formulation \emph{all}) subsets of the auctioned goods -- in our case, channels. It is important to emphasize here that in contrast to the previously mentioned approaches, which are based on the evaluation of single channels, this approach already considers the explicit evaluation of bundles of channels by tenants.

Following \cite{de2003combinatorial}, we formulate the combinatorial auction problem as follows. The bid announced by player (tenant) $k$ for the bundle  $S$ of channels is denoted by $b_k(S)$. In our case, this value will be determined either by $\rho_k(S)$ or $U_k(\rho_k(S))$, depending on the actual context (capacity or utility). $y(S,k)$ denotes the acceptance indicator of the bundle $S$ for the participant $k$, i.e. $y(S,k)=1$ if the bundle $S$ is assigned to player $k$, and 0 otherwise. Let us now consider the integer optimization problem described by eq. (\ref{CA_basic_form}).
\begin{align}
    & \text{max}\sum_{k} \sum_{S \subseteq CH}b_k(S)y(S,k) \nonumber \\
    & \text{~s.t.~} \sum_{S \ni m} \sum_{k} y(S,k) \leq 1~~\forall m \in CH,  \qquad \sum_{S \subseteq CH} y(S,k) \leq 1 ~~~\forall k \label{CA_basic_form}
\end{align}

The first constraint of eq. (\ref{CA_basic_form}) ensures that overlapping goods are never assigned, while the second constraint ensures that no bidder receives more than one bundle.

CA-based methods have been already proposed for telecommunication resource sharing problems. 
While \cite{Pal2007} uses the CA principle for a scheduling problem, \cite{Xu2012} describes a reverse combinatorial auction game to allocate downlink resources in a cellular environment. Finally \cite{Wang2015} uses CA in a device-to-device (D2D) setup, where channels and power are jointly allocated.

In \cite{hasan2015distributed}, three approaches are proposed for distributed resource allocation utilizing the concepts of stable matching, factor graph-based message passing, and distributed auction in order to maximize the system throughput. No numerical results are presented, however, a qualitative comparison is performed.

For every subset $S$ of $CH$, $b_k(S)$ denotes the value, which the given subset represents to tenant $k$. This value is either equal to the rate $\rho_k(S)$ ensured to tenant $k$ by the bundle of channels in $S$, or to the utility value ensured by this rate $U_k(\rho_k(S))$.

Regarding the application of CA for channel allocation in multi-connective environments, even with a relative small number of channels (e.g. 10), the cardinality of the set containing all subsets of $CH$ is very high, which makes the execution of the algorithm infeasible. On the the hand, if the channel number allows, the CA problem described in eq. (\ref{CA_basic_form}) boils down to brute-force optimization and determines the global optima of (\ref{eq_assignment_opt}).

In the general case, when the number of channels does not allow the 'full' optimization via CA, our proposed approach is the \emph{preallocation} of channels to tenats. Following this preallocation, tenants will formulate bids for the possible subsets of the channels preallocated to it, significantly decreasing the number of bids, if the cardinality of the preallocated channels is low enough. The set of channels relevant for tenant $k$ will be denoted by $CH^k$.
Let us emphasize that this preallocation of channels is not exclusive in the sense, that typically a channel is preallocated to multiple tenants during the preallocation process (otherwise the CA would make no sense, and every tenant would receive the full bundle of preallocated channels, since nobody else would bid on them).

\subsubsection{Preallocation of channels}
\label{subsec_preallocation}

This \emph{preallocation} of channels is carried out in the proposed method via the many-to-may version of the GS, where the preferences are set up based on the single-connectivity capacity/utility values.
In this version of the GS, both the channels and the tenants are characterized by a quota, denoted respectively by $q_{ch}$ and $q_{t}$.
In the first step of the process, each channel proposes to the most preferred $q_{ch}$ tenant, and each tenant puts the most preferred $q_{t}$ proposals on hold, and rejects the rest. The following steps are the same as in the case of the basic GS. The algorithm stops, if there are no rejected proposals. In the final state, most of the channels will be allocated to multiple tenants. Sets of channels allocated to tenants constitute the $CH^k$ sets, which are potentially overlapping.

Let us note that as the many-to-many version of the GS does not guarantee in general that every channel will be preallocated (this depends on the exact value of the quotas of channels and tenants). In such cases, we preallocate the remaining channels at random (to multiple tenants) to avoid the presence of unused resources.
In the current study we performed this step considering the constraint, that every tenant may have at most 8 preallocated channels. This constraint was based on the consideration that a high number of preallocated channels implies a very high number of bids in the CA method (the number of submitted bids grows exponentially with the number of preallocated channels).

%Note that if the sets $CH^k$ contain all available channels, it boils down brute force optimization and the global optimum can be determined. (maybe a numerical comparison for small size problem instance). 

\subsubsection{Determination of bids}

We will represent the set of the bids submitted by tenant $k$ with a matrix of $|CH|+2$ columns and $n_B$ rows, called the \emph{bid matrix}, where $n^k_B$ is the total number of bids submitted by tenant $k$.
The first $|CH|$ columns of the matrix corresponds to the individual channels. These columns can hold ones or zeros, depending on whether the actual channel is included in the bundle of channels corresponding to the actual bid or not (rows correspond to bids). The $|CH|+1$-th column corresponds to the value of the bid, which is equal to the capacity or utility value implied by the actual subset of channels, while the last column indicates the player (tenant) who submitted the bid.
A simple example demonstrating the structure of the bid matrix may be found in Appendix B.

\subsubsection{Basic combinatorial auction method (\textbf{CA})}

The combinatorial auction algorithm is executed as a linear integer optimization problem, where the variables are the (binary) acceptance indicators of single bids.
The objective is to maximize the total value of the accepted bids, while the constraints describe the following two considerations: (1) For each bidder, maximum one bundle may be assigned, and (2) Each item (channel) may be assigned to maximum one bidder.

\subsubsection{Fairness-enhanced combinatorial auction method (\textbf{FECA})}
\label{subsec_FECA}

The standard CA method optimizes the total value of accepted bids, but does not guarantee any minimum value for any participant.
To make the method more fair, the CA formulation may be straightforwardly extended to include (linear) constraints, which describe a minimum resulting value for each player. In the current study, in case of capacity context, we assume that this minimum value is equal to the $C_k^{min}$ parameter of the utility function for each player, while in the utility-based case, we assume that this required minimum value is equal to 1/3.
However, in contrast to the basic CA method, which always has a feasible solution (e.g. the trivial solution, when no channels are assigned to any tenant), in the case of FECA, it is possible that adding the minimum value constraints for the participants makes the optimization problem infeasible. In this case, we iteratively decrease the required minimum values for all tenants by a factor of 0.5, until the problem becomes feasible.

% proportional fairness scheduling: use $U_k(x) = log(x)$
% http://www.statslab.cam.ac.uk/~frank/elastic.pdf

\section{Results}
\label{sec_results}

\subsection{Simulation setup}
\label{subsec_simulation_setup}
We considered a 100x50 m rectangular area, with 8 BSs located on the walls at random positions.
Regarding the connectivity model described in subsection \ref{subsec_connectivity_model} , the invariant parameters were the following. The channel bandwidth ($B$) was considered with the value of 20 MHz, the reference value for the
Rician Factor ($K_{ref}$) was  14.1 dB, while the reference distance ($d_0$) was assumed to be equal to 15m. The interference power ($P_I$) has been taken into account with the value of -50 dBm, while the reference path loss ($PL(d_0)$) was 70.28 dB. The path loss exponent ($\delta$) was 2. As mentioned earlier, we assumed an outage probability threshold $\varepsilon$= $10^{-9}$.
    
The transmit power ($P_T$) of each BS was randomly chosen from the interval [15 25] dBm, assuming uniform distribution. % heterogeneous versus homogenous BS 
Each BS offers 1, 2, or 3 channels (with equal probability), but the total number of channels was limited to 20. 

% cell-less or cell-free deployment "Prospective Multiple Antenna Technologies for Beyond 5G" 
% Put in context of B5G and 6G - todo EJ

We assumed 6 tenants, located at random positions. In order to model heterogeneous service requirements of the tenants, the $C^{min}$ parameter has been chosen randomly from the interval [0.1 0.2], while the parameter $C^{max}$ was chosen randomly from the interval [15 25] both according to uniform distribution. 

In the case of the R, SR1 and SR2 methods we assumed 
$\bar{n}_{ch}=4$.
In the case of GS, we assumed a quota of 4 for each tenant, while in the case of the CA and FECA methods, we assumed a quota of 6 for both the tenants and the channels during the preallocation process. The minimal capacity defined in the FECA was equal to $C_min$ for each tenant in the capacity context and the minimal utility was set to $1/3$ in the utility context.

To test the robustness of the analyzed allocation algorithms, we considered an obstacle-free case (case I) as reference, when $K_{k,i}=K_{ref} ~\forall~(k,i)$, and two perturbed cases, in which we assume that obstacles are negatively affecting the connectivity between certain BS-tenant pairs (in other words, we assume channel outages). In case II, the $K_{k,i}$ values have been reduced to 0 for the 25\% of the possible BS-tenant pairs, while in case III, the $K_{k,i}$ values have been reduced to 0 for the 50\% of the possible BS-tenant pairs.
The exact pairs, for which the $K_{k,i}$ values have been reduced have been chosen at random in each scenario.

10000 scenarios have been simulated. For the fundamentally non-deterministic  methods (R, SR1 and SR2), 10 runs have been performed for each scenario, and the average resulting values have been considered as results for each scenario. The simulations have been performed on a desktop computer, with an Intel core i5 processor @2.9 GHz and 16GB of RAM, using 64 bit Windows and MATLAB.

%=======================
%\newpage
\subsection{Total throughput / total utility}

\begin{figure}
     \centering
     \begin{subfigure}[b]{0.48\textwidth}
         \centering
         \includegraphics[width=\textwidth]{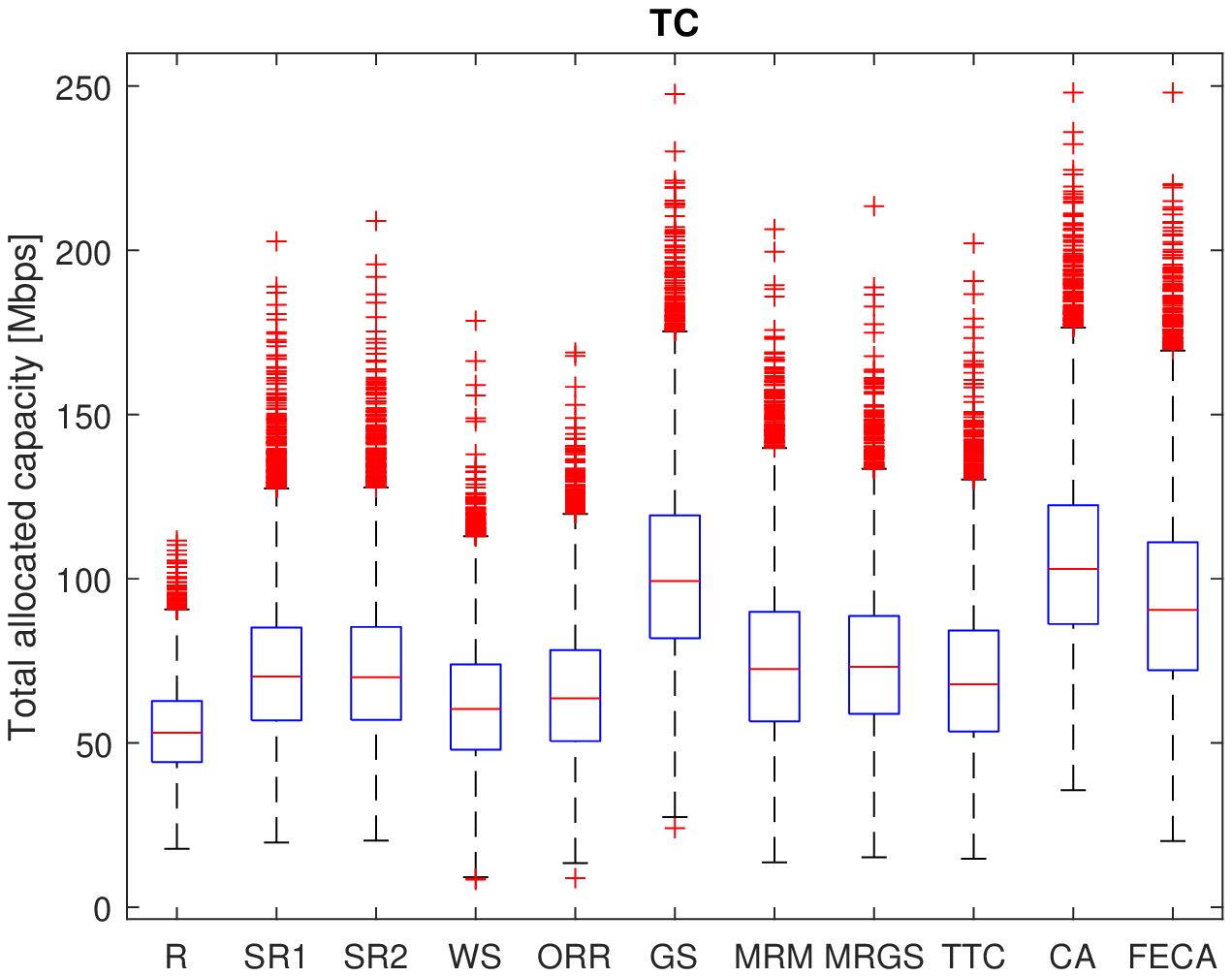}
         \caption{}
         \label{Fig_capacity_TC_0red}
     \end{subfigure}
     \hfill
     \begin{subfigure}[b]{0.48\textwidth}
         \centering
         \includegraphics[width=\textwidth]{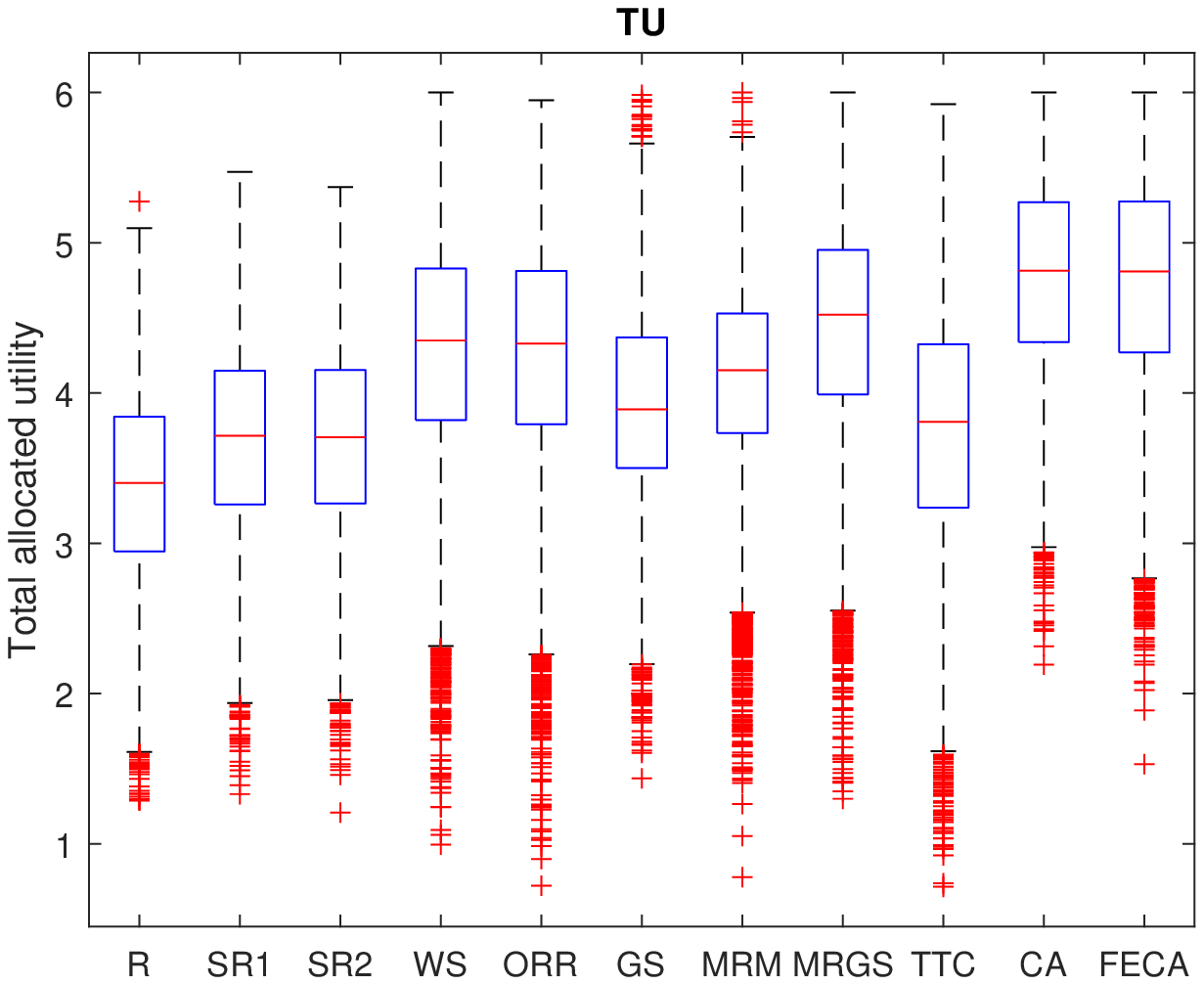}
         \caption{}
         \label{Fig_utility_TU_0red}
     \end{subfigure}
     \hfill
     \begin{subfigure}[b]{0.48\textwidth}
         \centering
         \includegraphics[width=\textwidth]{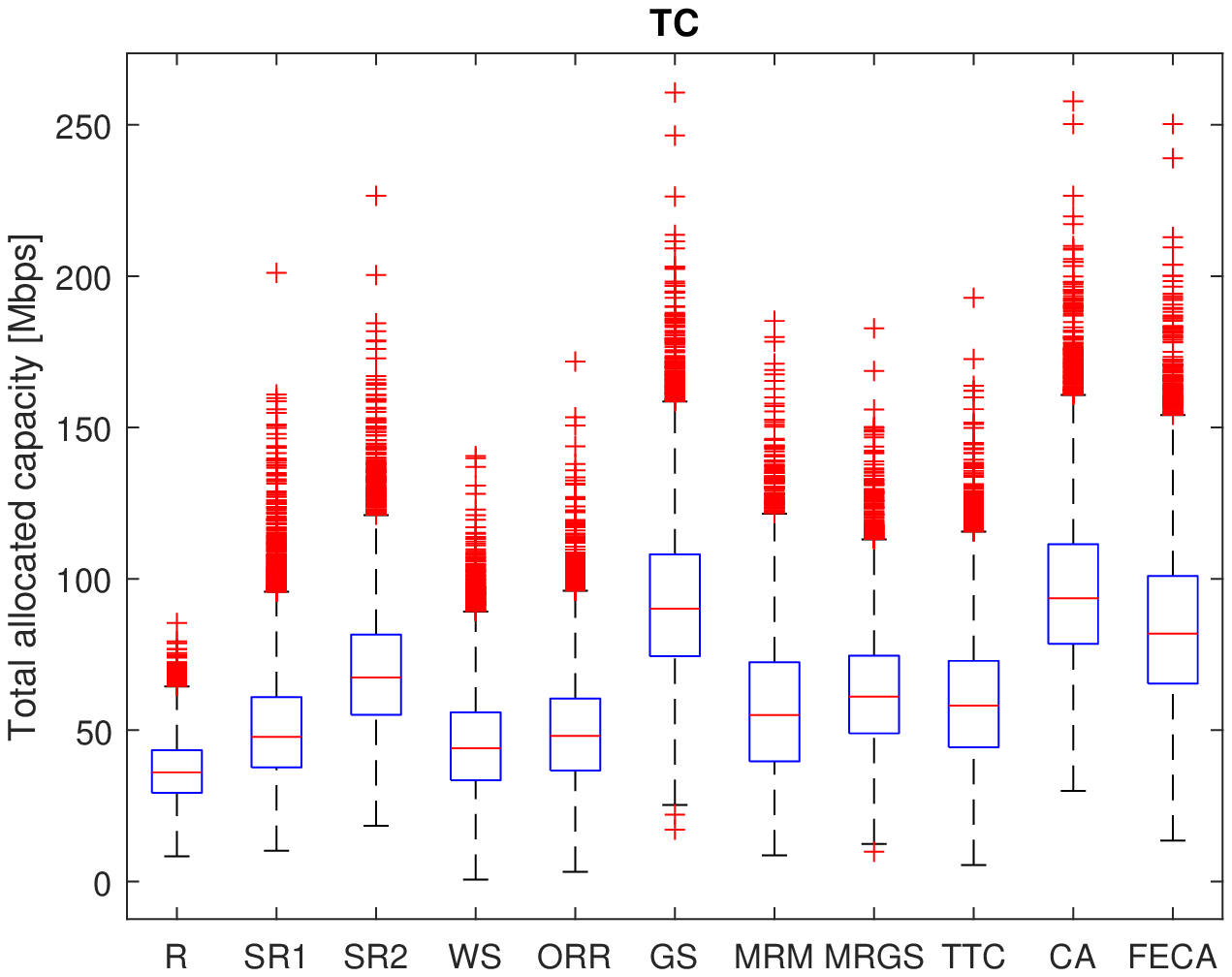}
         \caption{}
         \label{Fig_capacity_TC_25red}
     \end{subfigure}
     \hfill
     \begin{subfigure}[b]{0.48\textwidth}
         \centering
         \includegraphics[width=\textwidth]{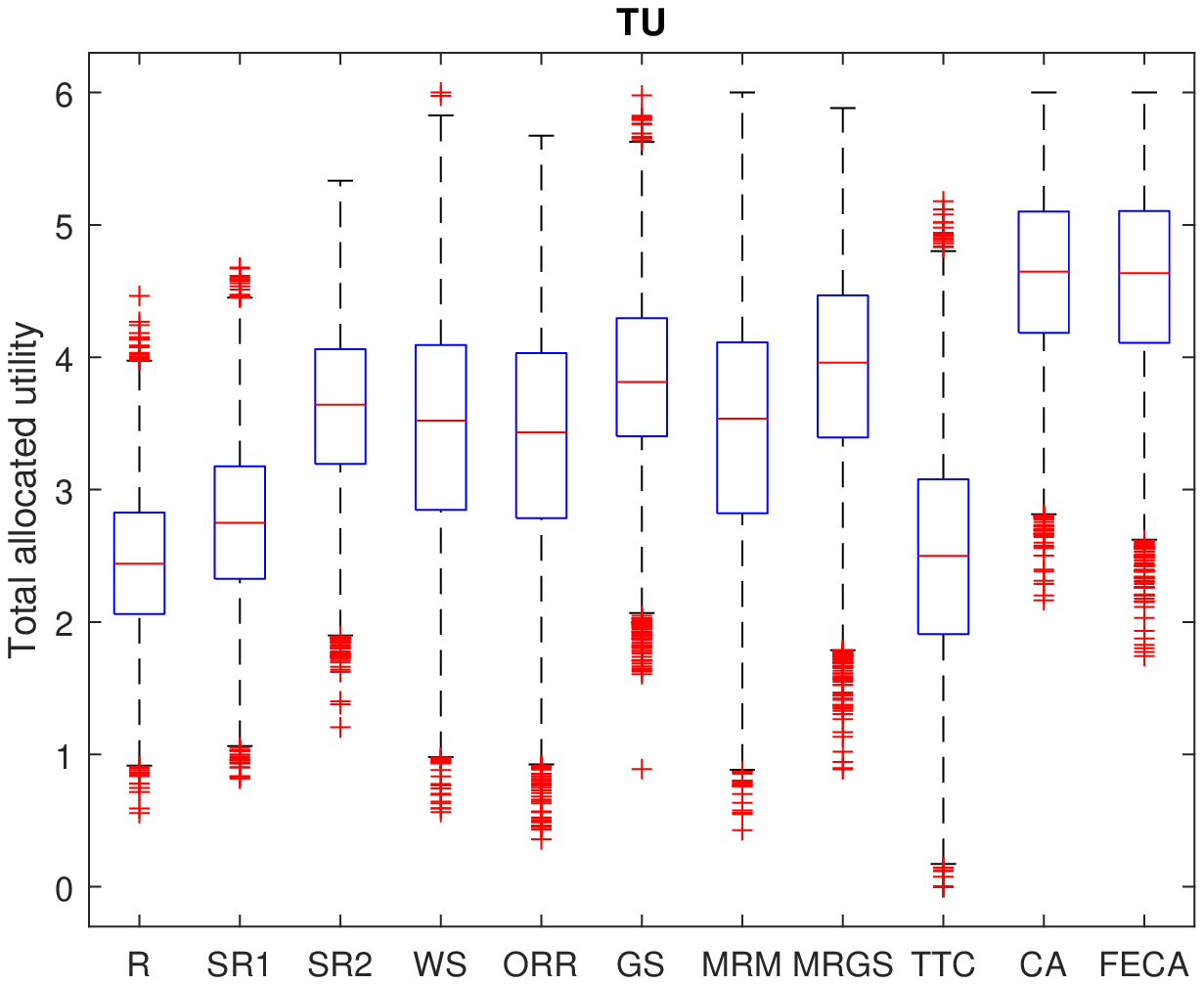}
         \caption{}
         \label{Fig_utility_TU_25red}
     \end{subfigure}
     \hfill     
     \begin{subfigure}[b]{0.48\textwidth}
         \centering
         \includegraphics[width=\textwidth]{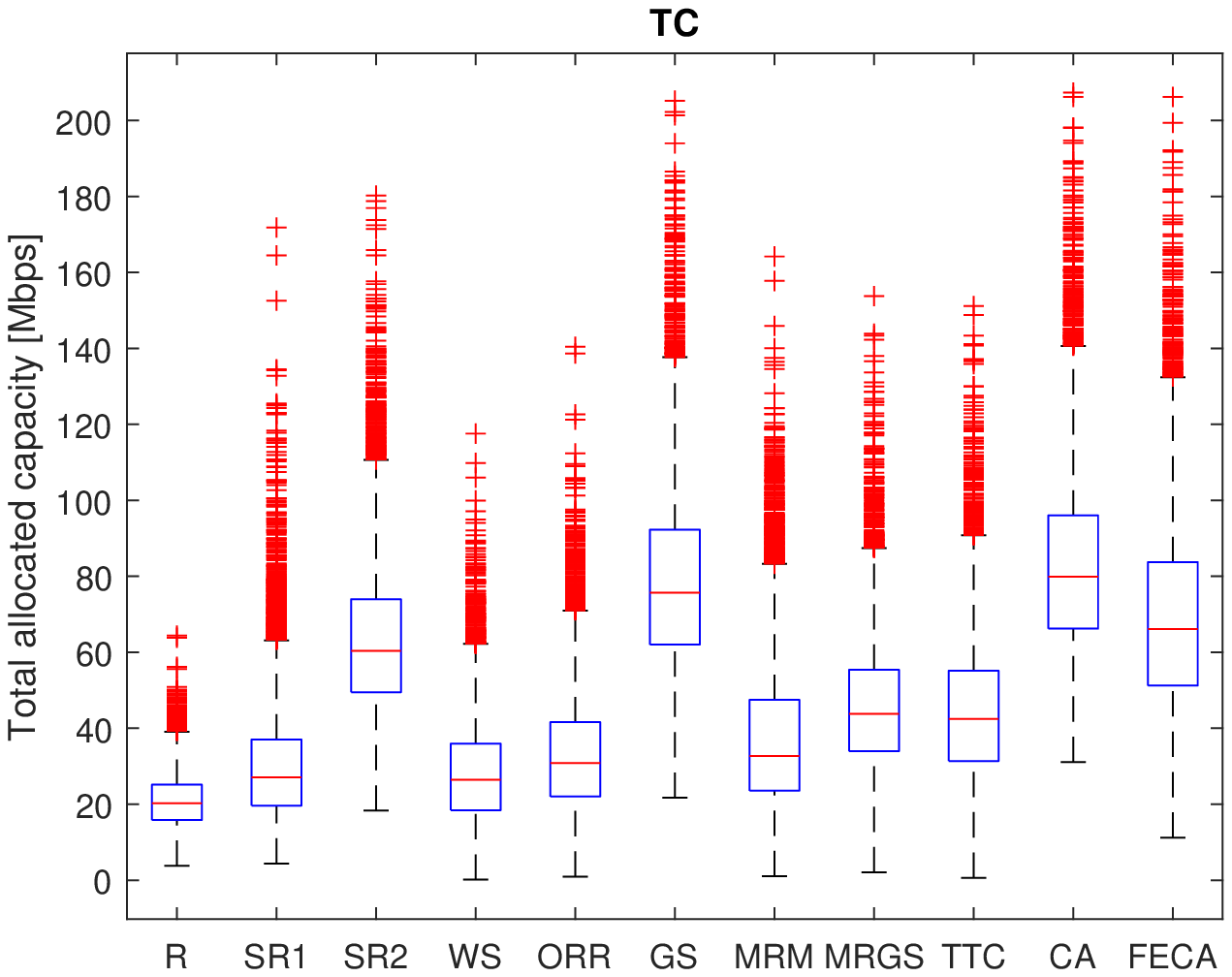}
         \caption{}
         \label{Fig_capacity_TC_50red}
     \end{subfigure}
     \hfill
     \begin{subfigure}[b]{0.48\textwidth}
         \centering
         \includegraphics[width=\textwidth]{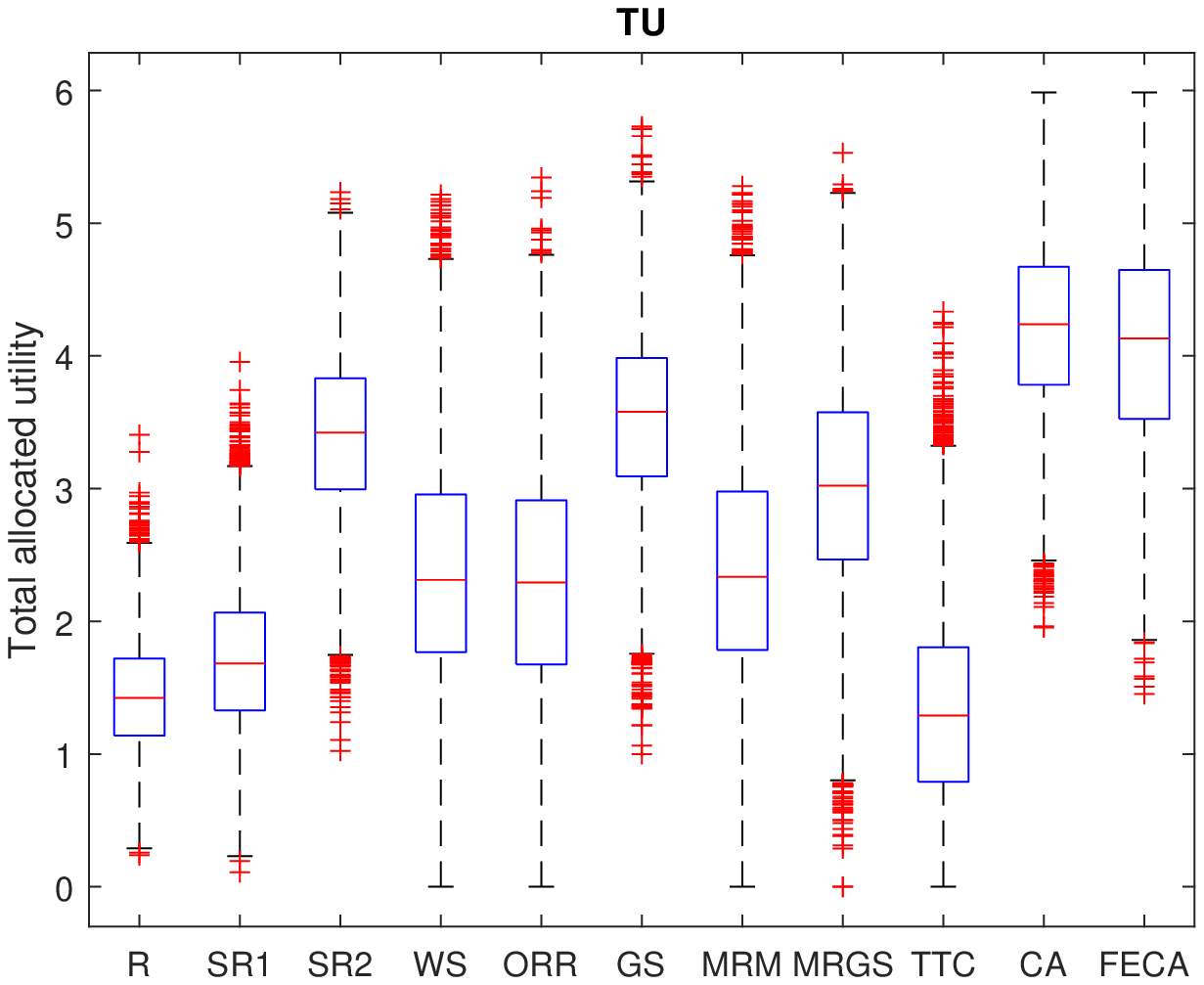}
         \caption{}
         \label{Fig_utility_TU_50red}
     \end{subfigure}     
     \hfill
     \caption{Overall total performance of the system in capacity (a,c,e) and utility (b,d,f) context in cases I (a,b), II (c,d) and III (e,f). \label{Fig_TCTU}}
\end{figure}

The most straightforward measure of overall system performance is the total resulting value of the quantity, which we aim to maximize, namely the total allocated capacity ($TC$) -- or total throughput -- in the capacity context, or the total resulting utility ($TU$) in the utility context.
Figure \ref{Fig_TCTU} depict the results in cases I , II and III in the capacity and in the utility context.
Figures \ref{Fig_capacity_TC_0red} - \ref{Fig_utility_TU_50red}
clearly show that as the resources become more scarce, the performance of the algorithms is (not surprisingly) decreased.
In the box plots, the central mark is the median, while the edges of the box are the 25th and 75th percentiles respectively. The whiskers extend to the most extreme data points which are considered not to be outliers, and the outliers are plotted individually with red crosses.
Tables \ref{Tab_capacity_TC} and \ref{Tab_utility_TU} in Appendix C summarize the median and mean values of total allocated capacity and utility in the case of various allocation methods.

%=======================
%\newpage
\subsection{Fairness}

The second important aspect is the fairness of the resulting allocation.

%however it is not defined in the problem formulation

Fairness may be measured through different quantities. Here, as a primary measure for fairness, we used the quantities defined as follows
\begin{align}
& F_C(A)=\prod_{i=1}^{n_T}C_k, \label{eq_fairness_measure_capacity} \qquad F_U(A)=\prod_{i=1}^{n_T}U_k(C_k) \end{align}
We interpret $F_C(A)$ and $F_U(A)$ as dimensionless quantities.

\begin{figure}
     \centering
     \begin{subfigure}[b]{0.48\textwidth}
         \centering
         \includegraphics[width=\textwidth]{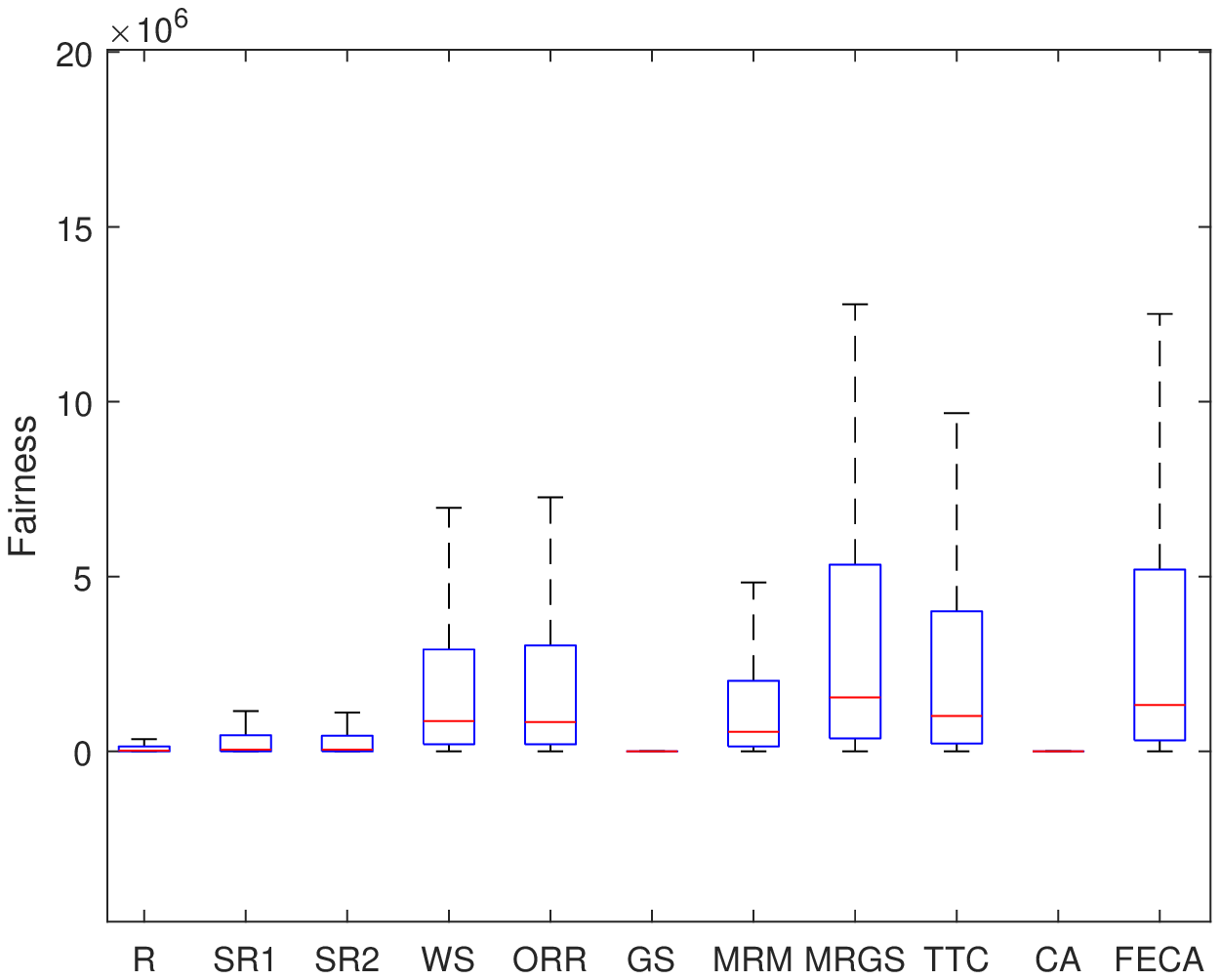}
         \caption{}
         \label{Fig_capacity_F_0red}
     \end{subfigure}
     \hfill
     \begin{subfigure}[b]{0.48\textwidth}
         \centering
         \includegraphics[width=\textwidth]{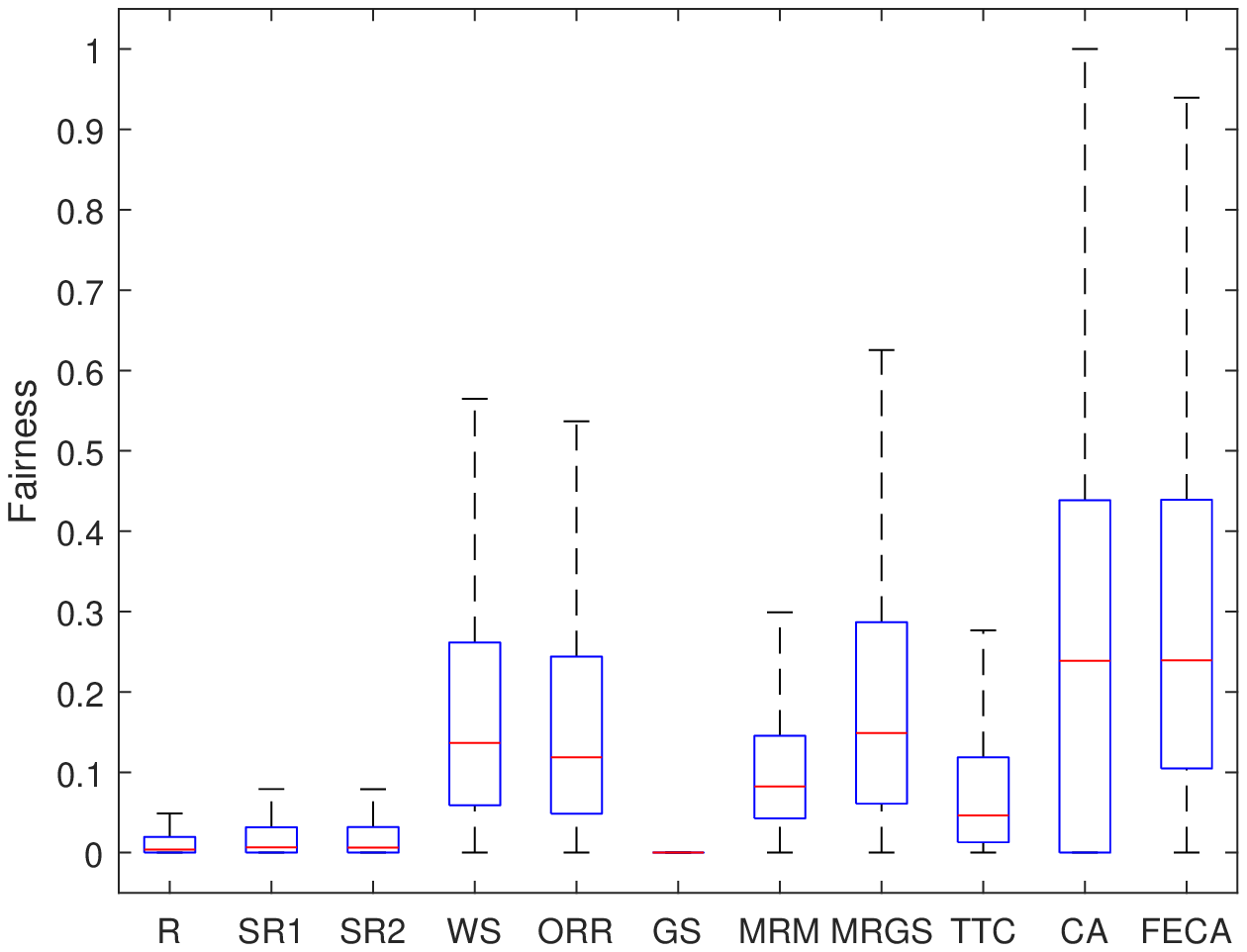}
         \caption{}
         \label{Fig_utility_F_0red}
     \end{subfigure}
     \hfill
     \begin{subfigure}[b]{0.48\textwidth}
         \centering
         \includegraphics[width=\textwidth]{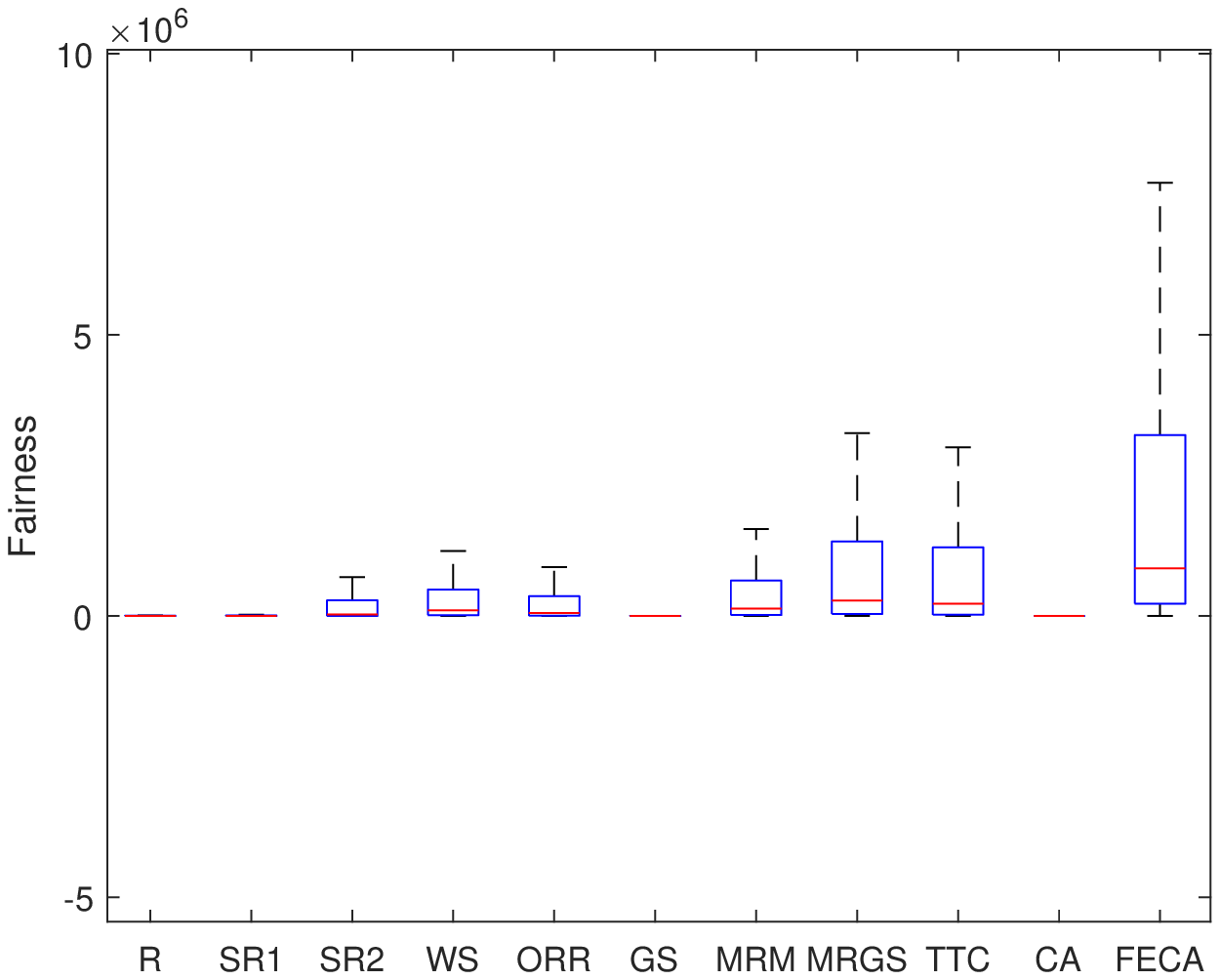}
         \caption{}
         \label{Fig_capacity_F_25red}
     \end{subfigure}
     \hfill
     \begin{subfigure}[b]{0.48\textwidth}
         \centering
         \includegraphics[width=\textwidth]{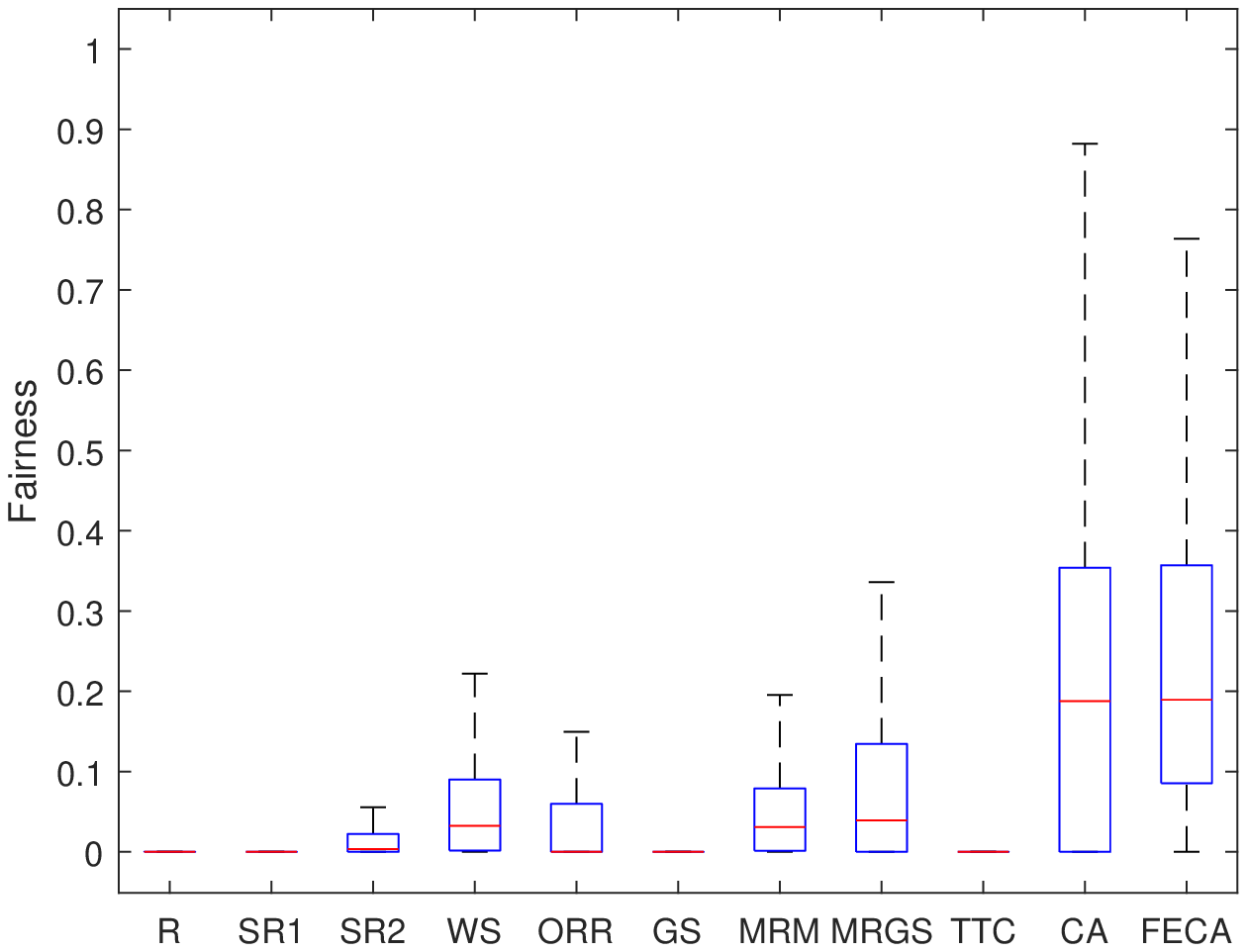}
         \caption{}
         \label{Fig_utility_F_25red}
     \end{subfigure}
     \hfill     
     \begin{subfigure}[b]{0.48\textwidth}
         \centering
         \includegraphics[width=\textwidth]{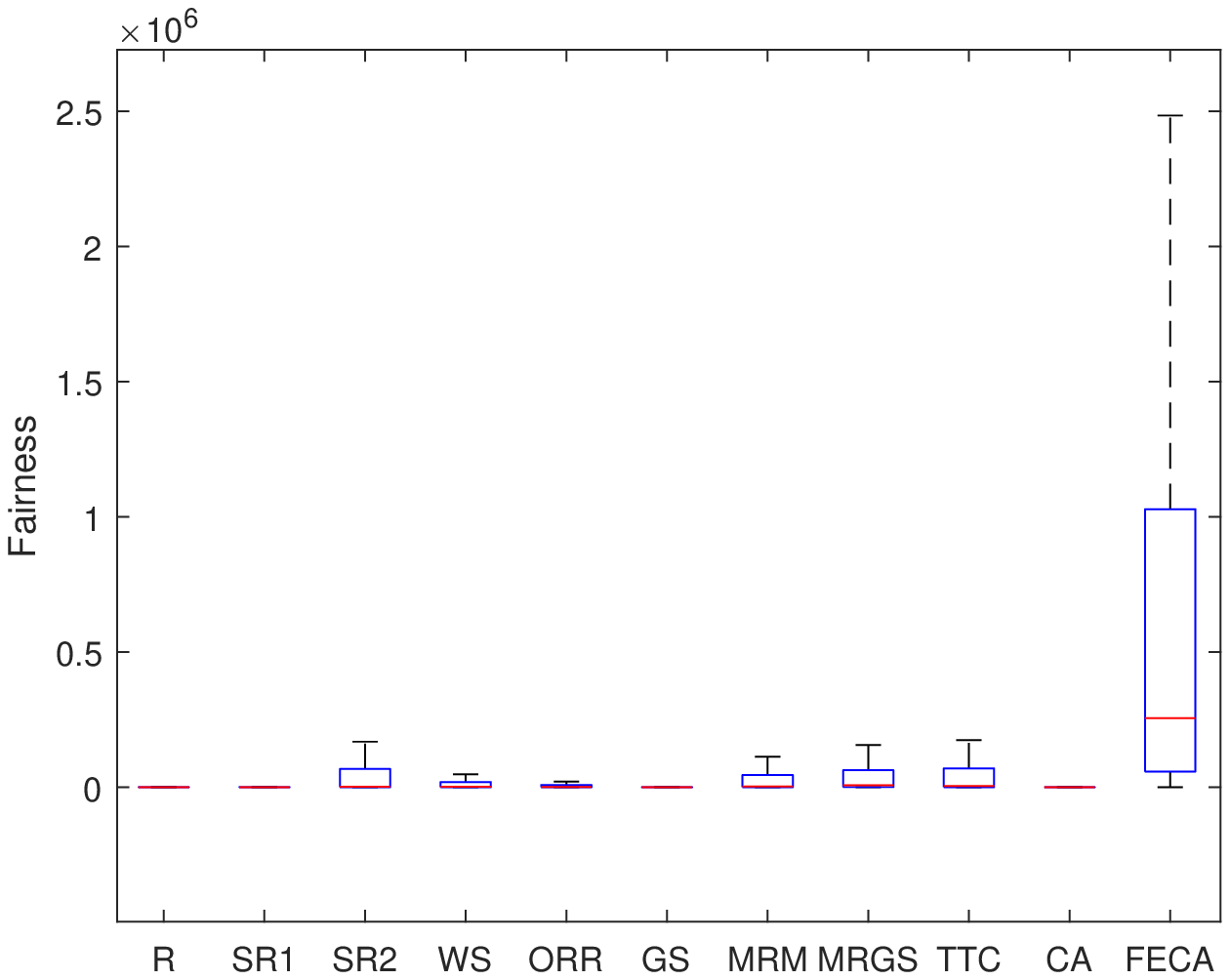}
         \caption{}
         \label{Fig_capacity_F_50red}
     \end{subfigure}
     \hfill
     \begin{subfigure}[b]{0.48\textwidth}
         \centering
         \includegraphics[width=\textwidth]{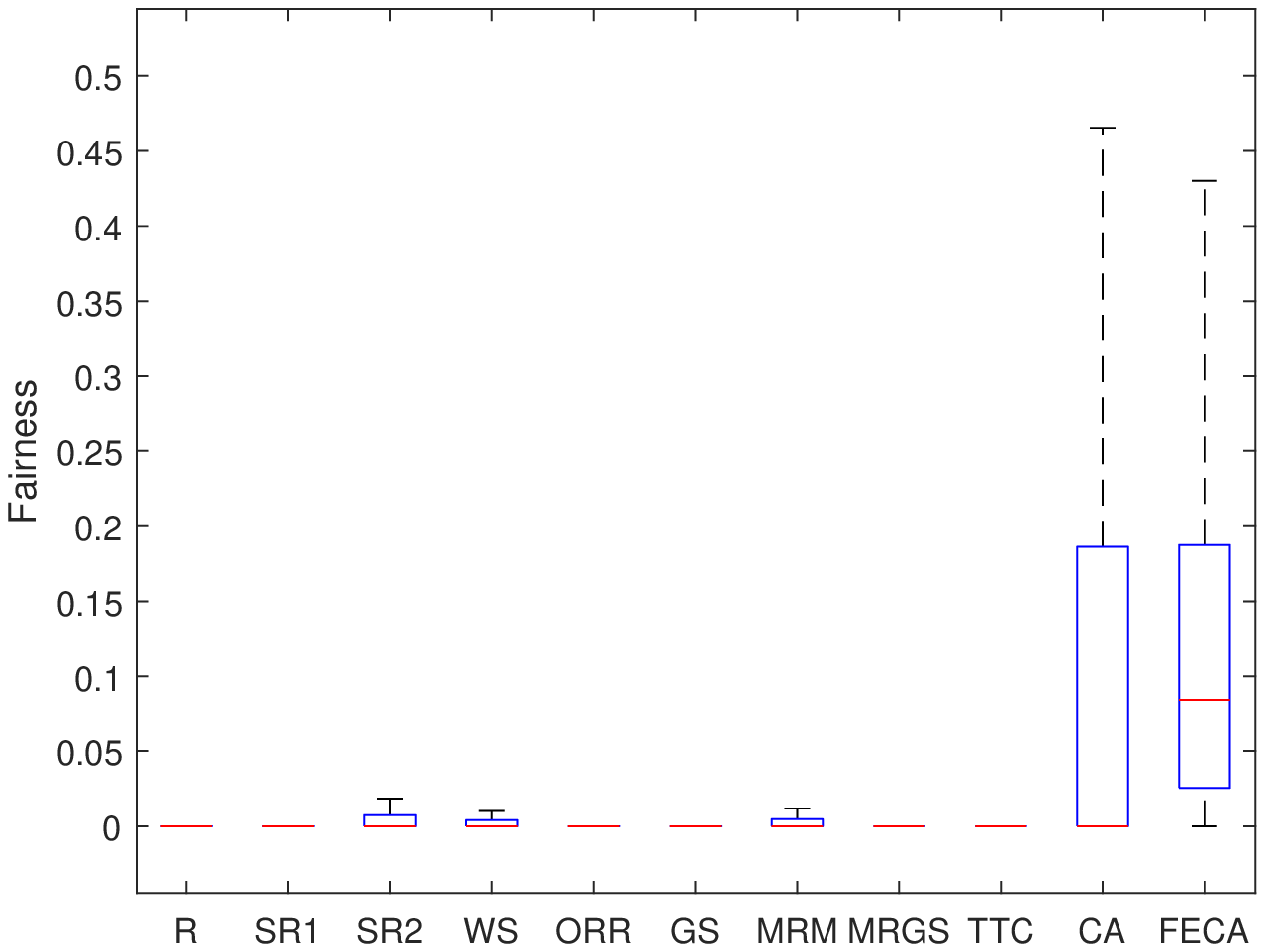}
         \caption{}
         \label{Fig_utility_F_50red}
     \end{subfigure}     
     \hfill
     \caption{Resulting values of the $F_C$ and $F_U$ in the case of capacity (a,c,e) and utility (b,d,f) -based allocation in cases I (a,b), II (c,d) and III (e,f). Outliers are not depicted for better visibility. \label{Fig_F}}
\end{figure}

Figure \ref{Fig_F} depict the results in cases I , II and III in the capacity and in the utility context. Tables \ref{Tab_capacity_F} and \ref{Tab_utility_F} in Appendix C summarize the median and mean values of fairness results.

In addition to the fairness measure defined by eq. (\ref{eq_fairness_measure_capacity}), we may also look at the 
resulting capacity/utility values of 'worst case' tenants with minimal capacity/utility values. The average minimal capacity/utility values in the case of various allocation methods are summarized in table \ref{Tab_MCMU}. Table \ref{Tab_MCMU} clearly shows, that as resource scarcity arises (cases II and III), the minimal allocated capacity and utility values are also decreased.

%CsD: Bold typeface in tables for highest numbers

\begin{table}[h!]
    \centering
    \begin{tabular}{|c|c|c|c|c|c|c|c|c|c|c|c|c|}
    \hline
        & & R & SR1 & SR2 & WS & ORR & GS & MRM & MRGS & TTC & CA & FECA\\ \hline
\multirow{ 2}{*}{Case I} & mean($MC$) &  0.809 &    0.851 &    0.844 &    \textbf{7.407} &   5.861 &    0.278 &   3.560 &    5.850 &    5.431 &    0.038 &    3.357  \\ 
& mean($MU$) & 0.069 &    0.074 &    0.074 &    0.571 &    0.487 &    0.023 &    0.418 &    0.485 &   0.358 &   0.476 &    \textbf{0.584}  \\ \hline
\multirow{ 2}{*}{Case II} & mean($MC$) &  0.183 &    0.198 &    0.710 &    \textbf{4.504} &    2.465 &    0.330 &    2.746 &    3.360 &    3.392 &    0.067 &    3.373  \\ 
& mean($MU$) & 0.011 &   0.012 &    0.061 &    0.348 &    0.187 &    0.028 &    0.292 &    0.270 &    0.049 &    0.438 &    \textbf{0.557} 
  \\ \hline
\multirow{ 2}{*}{Case III} & mean($MC$) &  0.026 &   0.030 &    0.430 &    1.752 &    0.787 &    0.244 &    1.330 &    1.161 &    1.243 &    0.048 &    \textbf{2.894} 
  \\ 
& mean($MU$) & 0.000 &    0.001 &    0.035 &    0.100 &    0.028 &    0.019 &    0.096 &    0.069 &    0.001 &    0.262 &    \textbf{0.427}
  \\ \hline  
    \end{tabular}
        \hspace{1.ex}
    \caption{Average values of minimum allocated capacity ($MC$) and utility ($MU$) in cases I, II and III. $MC$ values correspond to capacity-based allocation, while the $MU$ values correspond to utility-based allocation. maximal values are emphasized with bold typeface.}
    \label{Tab_MCMU}
\end{table}

Figure \ref{Fig_MCMU} in Appendix C includes the box-plots depicting the resulting minimal capacity values in the case of varoius allocation methods.

As a further measure corresponding to the topic of fairness, the number of outage events may be regarded as an important indicator. In the current simulation framework, an outage event happens, if the resulting allocated capacity of a tenant is less than the required minimal capacity value ($C_k< C^{min}_k~\rightarrow~U_k(C_k)=0$). Table \ref{Tab_nOC} summarizes the number of outage events ($nOE$) for various allocation methods, and various levels of resource scarcity (Cases I, II and III). In this table, one may see that on the one hand, the $nOE$ increases in the case of resource scarcity (cases II and III) , and on the other hand, the values are practically the same, independent of the  nature of allocation (capacity-based or utility based).

% transform to probability (divide by 8)

\begin{table}[h!]
    \centering
    \begin{tabular}{|c|c|c|c|c|c|c|c|c|c|c|c|c|}
    \hline
        & & R & SR1 & SR2 & WS & ORR & GS & MRM & MRGS & TTC & CA & FECA\\ \hline
\multirow{ 2}{*}{Case I} & mean($nOE$) (CBA) &  1.170 &   1.209 &    1.207 &    0.035 &    0.034 &    1.576 &    0.029 &    0.034 &    0.041 &    2.076 &    0.127   \\ 
& mean($nOE$) (UBA)& 1.169 & 1.206 &    1.205 &    0.034 &    0.033 &    1.570 &    0.033 &    0.034 &    0.054 &    0.321 &    0.065  \\ \hline
\multirow{ 2}{*}{Case II} & mean($nOE$) (CBA)&  2.102 &   1.253 &    2.061 &    0.347 &    0.750 &    1.514 &    0.309 &    0.426 &    0.395 &    1.991 &    0.078   \\ 
& mean($nOE$) (UBA)& 2.118 &    1.265 &    2.078 &    0.345 &    0.776 &    1.524 &    0.348 &   0.450 &    1.568 &    0.346 &    0.051  
  \\ \hline
\multirow{ 2}{*}{Case III} & mean($nOE$) (CBA)&  3.373 &    1.464 &    3.210 &    1.583 &    1.984 &    1.666 &    1.274 &    1.289 &    1.227 &    2.040 &    0.142
  \\ 
& mean($nOE$) (UBA)& 3.376 &    1.472 &    3.228 &    1.336 &    2.004 &    1.680 &    1.317 &    1.325 &    3.379 &    0.621 &    0.140 
  \\ \hline  
    \end{tabular}
        \hspace{1.ex}
    \caption{Average number of outage events (nOE) in the case of capacity-based allocation (CBA) and utility-based allocation (UBA) for the various allocation methods in cases I, II and III.}
    \label{Tab_nOC}
\end{table}

Regarding Table \ref{Tab_nOC}, let us note that an outage event our case means that a tenant does not receive enough 'ultra reliable' capacity, in the sense, that ultra-reliability criteria has been already taken into account in the capacity calculations (see the eqs. in subsection \ref{subsec_used_conn_fncn}), assuming the probability threshold $\varepsilon=10^{-9}$ defined in subsection \ref{subsec_simulation_setup}. We deliberately used such a 'strict' definition of outage event to make the methods comparable according to this dimension as well.

% comment: We deliberately set up a simulation framework in which appropriate channel allocation is challenging, and the performance differences between various algorithms are well shown in the simulation results.

%=======================
%\newpage
\subsection{Computational time}

The required computational time ($t$) is also an important aspect of the applied algorithms. Table \ref{Tab_t} summarizes the required computational time of the algorithms in the case of the various setups. As this table shows, the computational time does not depend on the nature of allocation (capacity-based or utility based), and in most of the cases it is unaffected by resource scarcity as well. However in the case of CA and FECA, the computational time decreases as resources become more scarce. The explanation for this phenomena is the following. 
As described in subsection \ref{subsec_preallocation}, in the case of the CA-based methods, bids are generated for all possible subsets of the preallocated channels. If some of the preallocated channels are in outage for the tenant in question (the respective $K_{k,i}$ value is zero), the bid value for all of the combinations of these channels is zero. As the rows corresponding bids with 0 value (thus e.g. such bids) are removed from the bid matrix before the MILP calculation described in e.q. (\ref{CA_basic_form}), thus the number of variables is decreased. Naturally, as resources become more scarce, the chance for such bids is increased. As described in subsection \ref{subsec_preallocation}, in the current setup all tenants may have maximum 8 preallocated channels. As in case III 50 percent of the BS-tenant pairs are in outage, it may easily happen that one or more tenants have 4 (or more) preallocated channels providing 0 capacity. Furthermore, as $C^{min}$ must be exceeded to provide nonzero utility, these channels even combined with any working channel provide typically still 0 utility, thus 0 valued bids, which are removed from the bid matrix during the computations.
Figure \ref{Fig_t} in Appendix C includes the box-plots corresponding to these results.

\begin{table}[h!]
% update!
    \centering
    \begin{tabular}{|c|c|c|c|c|c|c|c|c|c|c|c|c|}
    \hline
        & & R & SR1 & SR2 & WS & ORR & GS & MRM & MRGS & TTC & CA & FECA\\ \hline
\multirow{ 2}{*}{Case I} & mean($t$) (CBA) &  0.001&    0.001 &    0.001 &    0.017 &    0.016 &    0.003 &    0.021 &    0.021 &    0.012 &    0.337 &    0.345
   \\ 
& mean($t$) (UBA)&  0.001 &    0.001 &    0.001 &    0.016 &    0.016 &    0.003 &    0.021 &    0.021 &    0.012 &    0.369 &    0.3667  \\ \hline
\multirow{ 2}{*}{Case II} & mean($t$) (CBA)& 0.001 & 0.001 &    0.001 &    0.018 &    0.016 &    0.003 &    0.020 &    0.021 &    0.012 &    0.303 &    0.307    \\ 
& mean($t$) (UBA)& 0.001 &    0.001  &    0.001 &    0.016 &    0.016 &    0.003 &    0.020 &    0.021 &    0.012 &     0.330 &    0.325  \\ \hline
\multirow{ 2}{*}{Case III} & mean($t$) (CBA)&  0.001 &    0.001 &    0.001 &    0.018 &    0.017 &    0.003 &    0.019 &    0.021 &    0.012 &    0.251 &    0.262  \\ 
& mean($t$) (UBA)& 0.001 &    0.001 &    0.001 &    0.016 &    0.016 &    0.003 &    0.018 &    0.021 &    0.012 &    0.264 &    0.270   \\ \hline  
    \end{tabular}
        \hspace{1.ex}
    \caption{Average computational time ($t$) in [s] in the case of capacity-based allocation (CBA) and utility-based allocation (UBA) for the various allocation methods in cases I, II and III.}
    \label{Tab_t}
\end{table}

%=================================

%\newpage
\section{Discussion}
\label{sec_discussion}

\subsection{Performance of random assignment methods}

While the purely random assignment (R) is used as a 'dummy' reference case and basis for comparison, the distance-based (SR1) and single-connectivity value-based (SR2) semi-random assignment methods are computationally still very simple methods with computational need of milliseconds in the analyzed scenarios (see table \ref{Tab_t}). It is remarkable that in some cases, SR2 shows comparable performance with other more sophisticated methods.
Figure \ref{Fig_TCTU}, clearly shows that if no resource scarcity is present (Case I), the performance of SR1 and SR2 is similar. In contrast, in cases II and III, the performance of SR2 is (not surprisingly) significantly better compared to SR1 (since even channels of close BSs may be in outage).
Comparing the capacity and utility contexts, one may notice that compared to other methods (like WS and ORR), SR1 (and in case I also SR 2) performs worse in the utility context as in the capacity-based allocation scenario. This may be explained by the fact that in the case of random assignment methods, the capacity requirements of tenants are not considered (only a maximum number of 4 channels is taken into account for each tenant). Let us note furthermore that SR2 performs surprisingly well in the terms of total throughput/utility in case III, under heavy resource scarcity. On the other hand, regarding fairness, R, SR1 and SR2 result in very low values (see figure \ref{Fig_F}).

\subsection{Performance of selection-based methods}
Selection based methods (WS, ORR and partially the MRM) allow tenants to choose from the available channels, without evaluating the actual channel from the perspective of other tenants or BSs.
This basic principle is the flaw of these algorithms. A channel which could provide very high capacity for one or more tenants may be easily assigned to any other tenant (for which it provides much less value), if it turns out to be its actual best choice.
As shown in figure \ref{Fig_F}, if resources are not scarce, this ensures relatively high values of $F_C$ and $F_U$ (since every badly supported tenant is always able to choose some channels appropriate for it).
These methods are performed in multiple rounds, and between rounds the potential gain of each (still available) channel is re-evaluated. As shown in figure \ref{Fig_utility_F_0red}, this results in high values of total utility, if no resource scarcity is present. Figure \ref{Fig_TCTU} however shows, that as resource scarcity arises, the performance of these algorithms degrades both in the capacity and in the utility context.
Figure \ref{Fig_MCMU} in Appendix C furthermore shows that the WS algorithm performs well in the context of average minimal capacity/utility values, especially in cases I and II.
As detailed in table \ref{Tab_t}, the average computational time of these algorithms compared to randomized methods is higher by one order of magnitude, but is still in the range of 10-20 ms in the case of the simulated scenarios.

\subsection{Performance of the GS and GS-based methods}

In accordance with literature results, which analyzed the GS method in the terms of total system throughput \cite{Hossler2019matching} 
% maybe other references here...
(in capacity context in our terminology), the performance of the GS method is very good compared to randomized or selection-based methods, and it requires very low computational effort (see table \ref{Tab_t}).
Furthermore, in the terms of total capacity/utility the GS performs very well, in almost all of the cases.
The only setup, where the GS underperforms compared to the WS, ORR, MRM and the MRGS is the Case I of the utility context.
The explanation for this phenomena is the following.

\begin{figure}[h!]
  \centering
  \includegraphics[width=.5\linewidth]{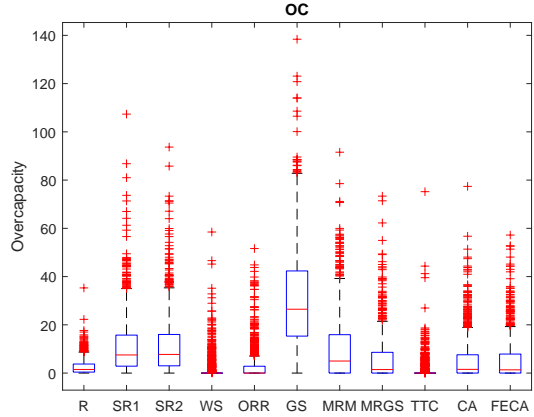}\\
  \caption{Overcapacity [Mbps] in Case I of the utility context} \label{Fig_utility_OC_0red}
\end{figure}

Selection-based algorithms and other GS-based algorithms are executed in a multi-round manner, and between the rounds, the single channels are re-evaluated according to already allocated utility values of tenants. This means that if a tenant already allocated enough capacity for itself (close to or over $C^{max}_k$), it wont aim for valuable channels anymore. In contrast, in the case of GS, since the many-to-one algorithm applies, all channels are allocated in a single round. If a tenant has good access to multiple (e.g. 5) valuable channels, they will be of course rated high in the preference list of the tenant, and if the tenant is the closest one to the BSs providing these channels, the tenant will be also rated high in the preference list of the channels. According to the principle of the GS, 4 of these channels will be allocated to the tenant (because of the quota of 4), even in the case if 3 of these channels (or maybe two of these channels combined with a weaker third) could provide the tenant with $C^{max}$ capacity, and thus already maximizing the utility.
This can be seen in Figure \ref{Fig_utility_OC_0red}, which clearly shows that the allocated overcapacity (capacity over $C^{max}$ is the highest in the case of the GS).

The performance of the GS strongly depends on the quota of the tenants. Typically there is an optimal quota for a setup (where the setup depends on $n_{ch}$, $n_T$), see \cite{Hossler2019matching}.  In the current paper we used the quota for the GS, which proved to be optimal in the simulated cases.

 In cases II and III, the number of useful channels is decreased, thus spectrum scarcity arises. As figure \ref{Fig_utility_OC_2550red} in Appendix C depicts, the trend is unaffected by the change, the average overcapacity is always the highest in the case of the GS. However, the value of the overcapacity is decreased, and as figures \ref{Fig_utility_TU_25red} and \ref{Fig_utility_TU_50red} show, the GS can provide a good performance in cases II and III also in the utility context despite of this drawback.
 
Figure \ref{Fig_F} clearly shows that while the GS is a good choice when one aims to optimize the total performance of the system, from the fairness perspective, the performance of the pure GS is poor. This can be seen also in Figure \ref{Fig_MCMU}, depicting the minimal allocated values of capacity and utility.

On the other hand, the MRM, and the MRGS which may be regarded as fairness-enhanced versions of the GS, provide good fairness results, while their total performance depicted in figure \ref{Fig_TCTU}  (especially of the MRGS) is higher compared to selection based methods (WS and ORR), which on the other hand provide also good fairness values (except for case III, where the resource scarcity becomes more serious, thus all methods struggle).

\subsection{Performance of the TTC}

According to our knowledge, the TTC algorithm has not been used in channel allocation problems. One aim of the current study was to analyze its performance in this framework. In fact, the results show that the TTC provides a relatively low performance compared to other methods (although it provides quite acceptable values of $F_C$ in the capacity context). Considering the principles of the algorithm, this is not surprising, since in TTC, a trade (an exchange of channels) occurs only if its beneficial to all of the involved participants. If e.g. a channel exchange between tenant 1 and tenant 2 would significantly enhance the rate of tenant 1 and slightly decrease the rate of tenant 2, it will not be performed in the TTC method.
Overall, the simulation results show that while the method may be applied in a computationally efficient way (see table \ref{Tab_t}) the application of TTC must be reconsidered in the multi-connectivity framework.

\subsection{Performance of the CA and FECA algorithms}

First of all, let us emphasize that as table \ref{Tab_t} also shows, the CA and FECA algorithms are the most computationally complex algorithms analyzed in this study. The required computational time, which is at least one magnitude higher compared to other methods (between 0.2 and 0.4 s) is however still in the range, which could be acceptable for setups with high performance instruments frequently used today in industry-automation applications. The sufficiency of these values naturally depend on the specific application. The computational needs of the CA and FECA strongly depend on the number of preallocated channels. In the current study this number was limited to 8, thus the number of submitted bids by participants is maximum $2^8-1$. As the possible number of preallocated channels is further increased, this value increases exponentially, thus the computational performance is quickly degraded. This points out one important aspect of the preallocation process: It must restrain the number of preallocated channels, to keep the required computational time at feasible levels.

In addition, in contrast to other methods, CA and FECA require the evaluation of channel bundles by tenants as well (while in the case of other methods, tenants are required only to evaluate single channels from time to time or once), which can mean a significant computational burden for tenants in some applications.

\begin{figure}
     \centering
     \begin{subfigure}[b]{0.48\textwidth}
         \centering
         \includegraphics[width=\textwidth]{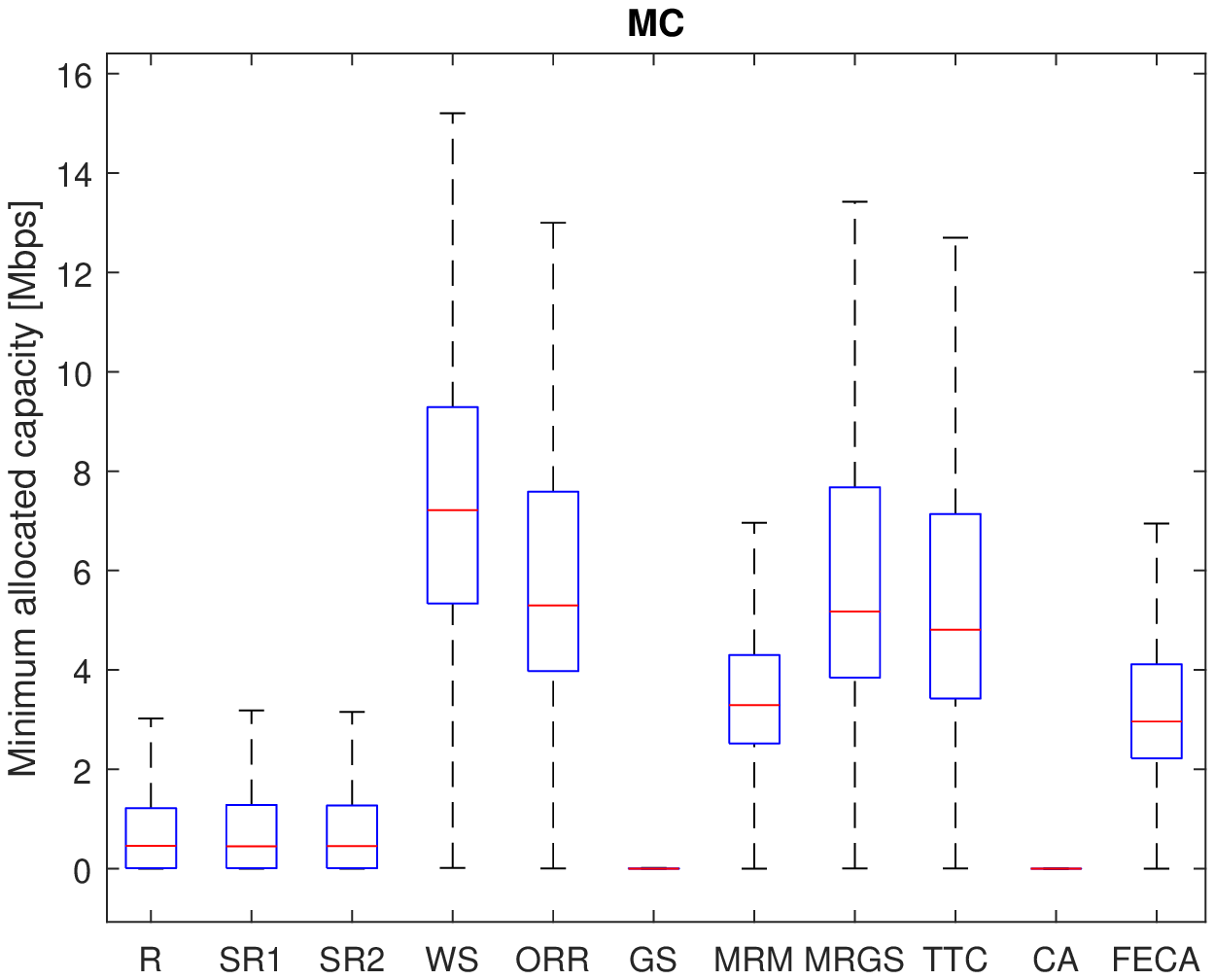}
         \caption{}
         \label{Fig_capacity_MC_0red}
     \end{subfigure}
     \hfill
     \begin{subfigure}[b]{0.48\textwidth}
         \centering
         \includegraphics[width=\textwidth]{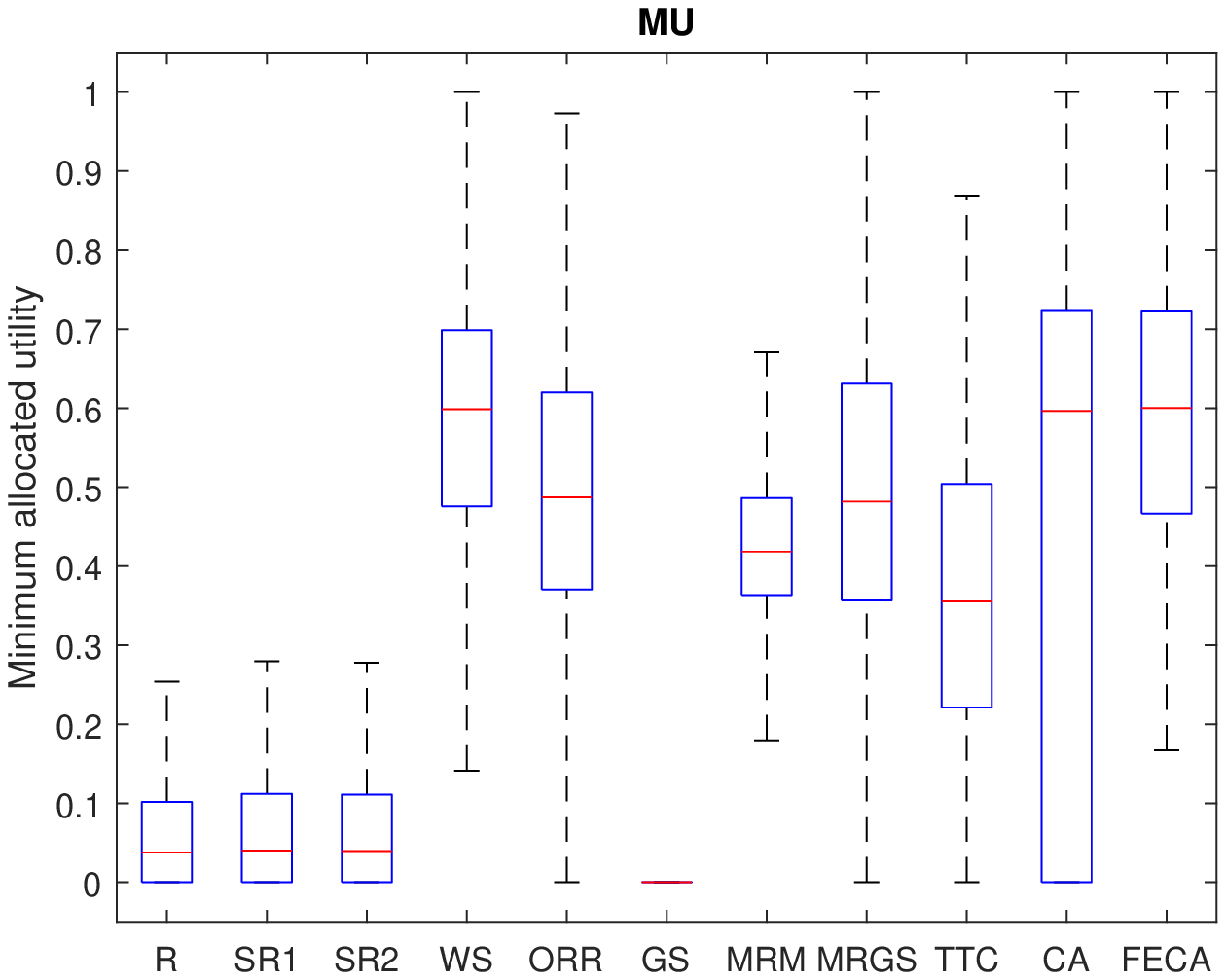}
         \caption{}
         \label{Fig_utility_MU_0red}
     \end{subfigure}
     \hfill
     \begin{subfigure}[b]{0.48\textwidth}
         \centering
         \includegraphics[width=\textwidth]{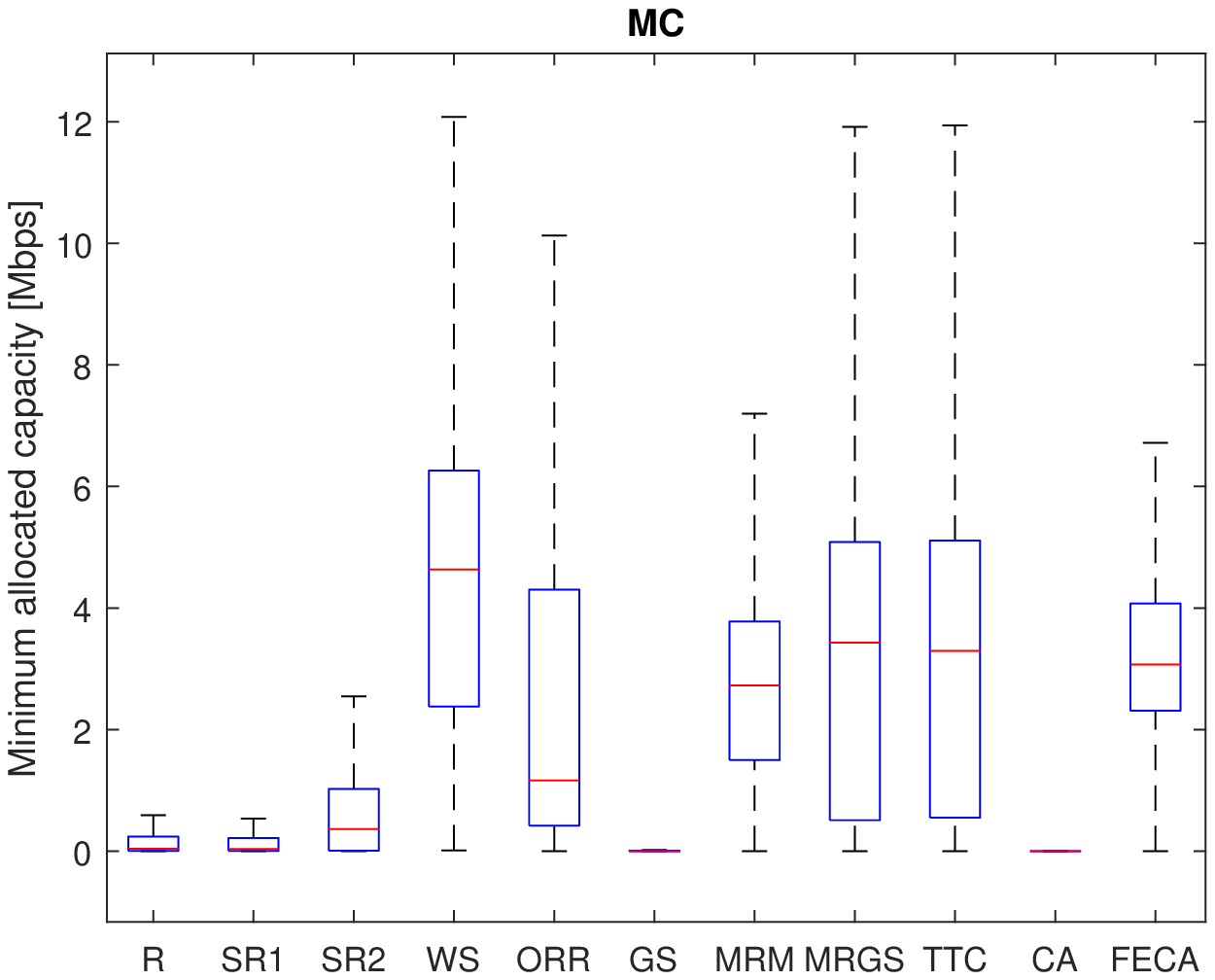}
         \caption{}
         \label{Fig_capacity_MC_25red}
     \end{subfigure}
     \hfill
     \begin{subfigure}[b]{0.48\textwidth}
         \centering
         \includegraphics[width=\textwidth]{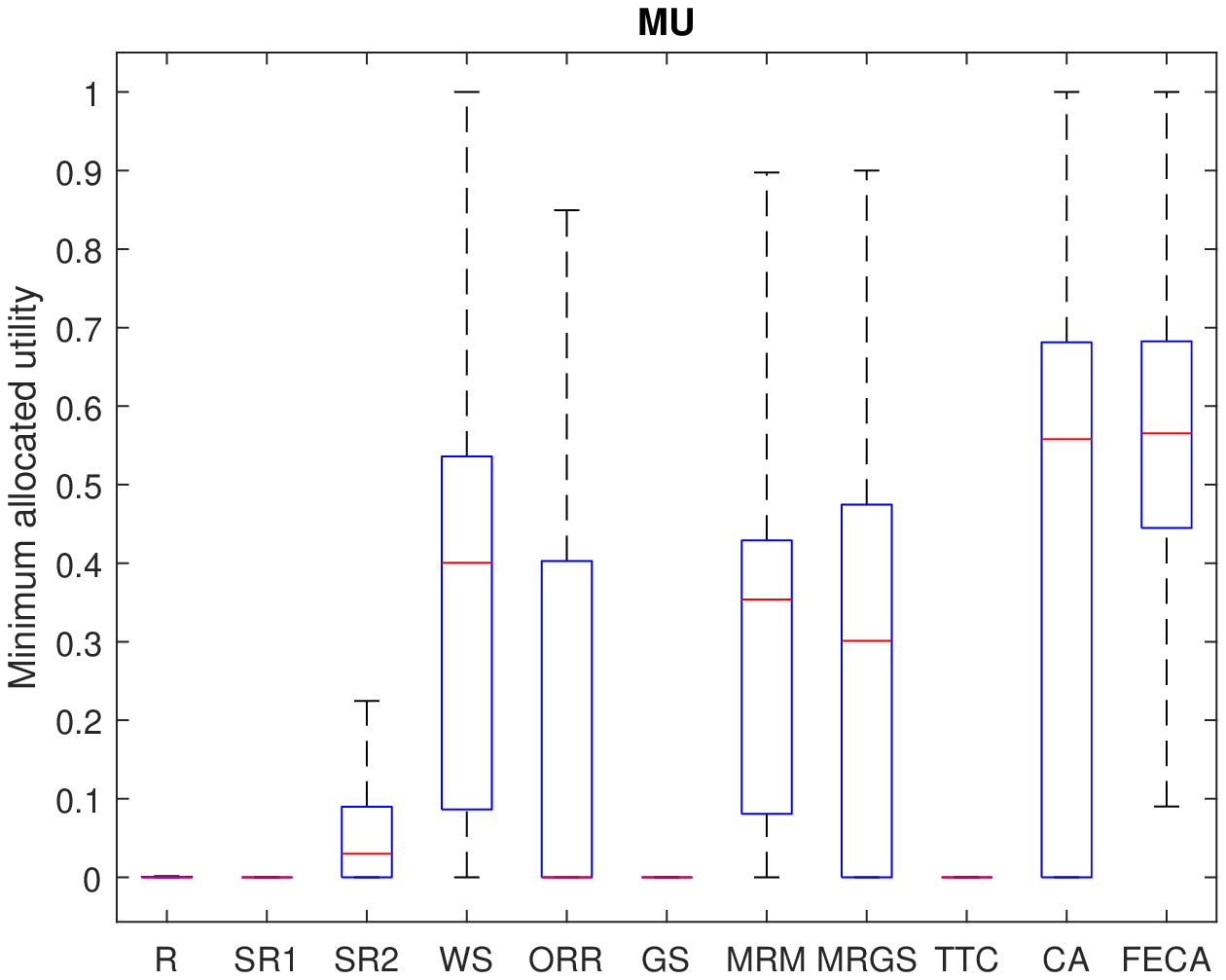}
         \caption{}
         \label{Fig_utility_MU_25red}
     \end{subfigure}
     \hfill     
     \begin{subfigure}[b]{0.48\textwidth}
         \centering
         \includegraphics[width=\textwidth]{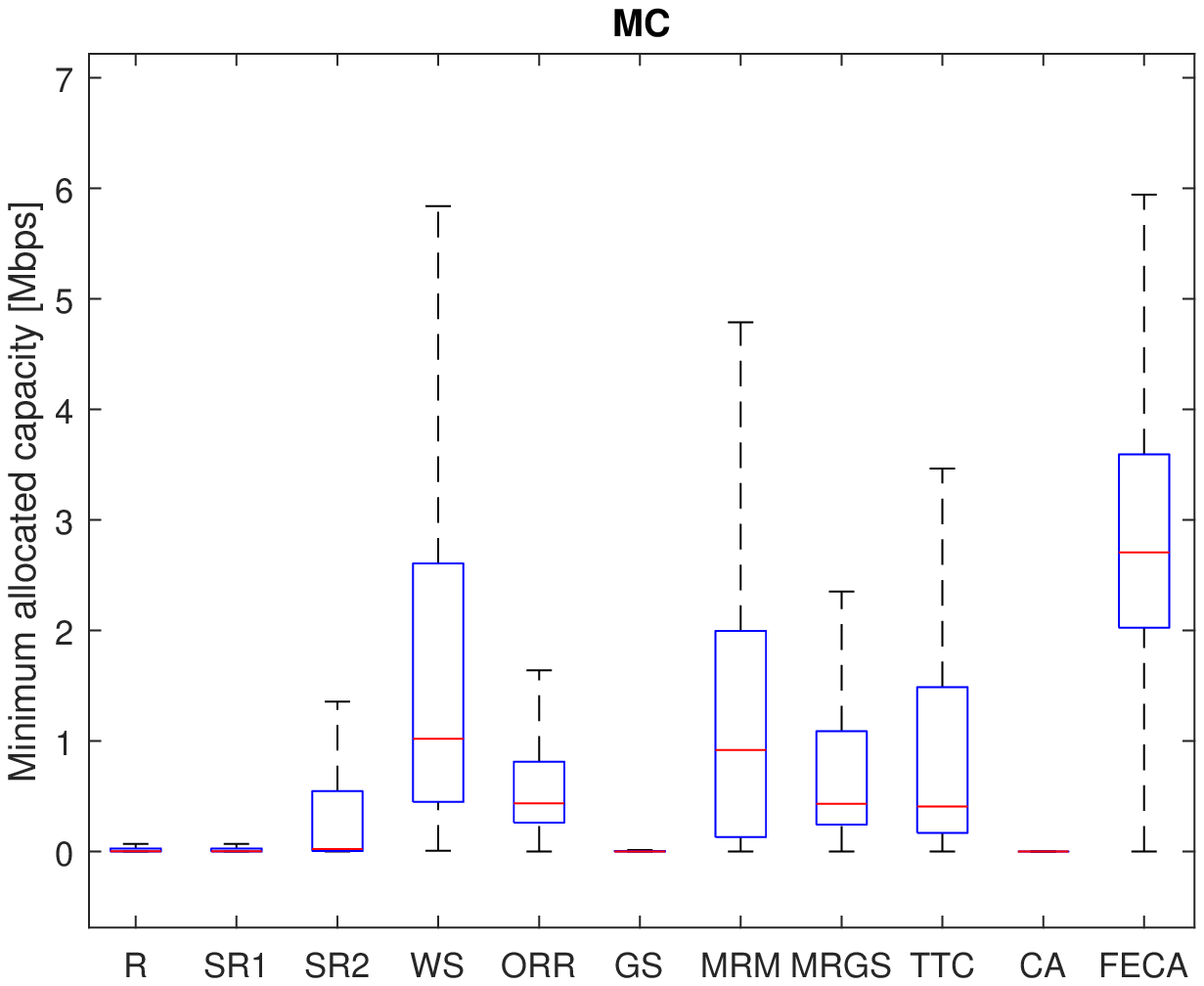}
         \caption{}
         \label{Fig_capacity_MC_50red}
     \end{subfigure}
     \hfill
     \begin{subfigure}[b]{0.48\textwidth}
         \centering
         \includegraphics[width=\textwidth]{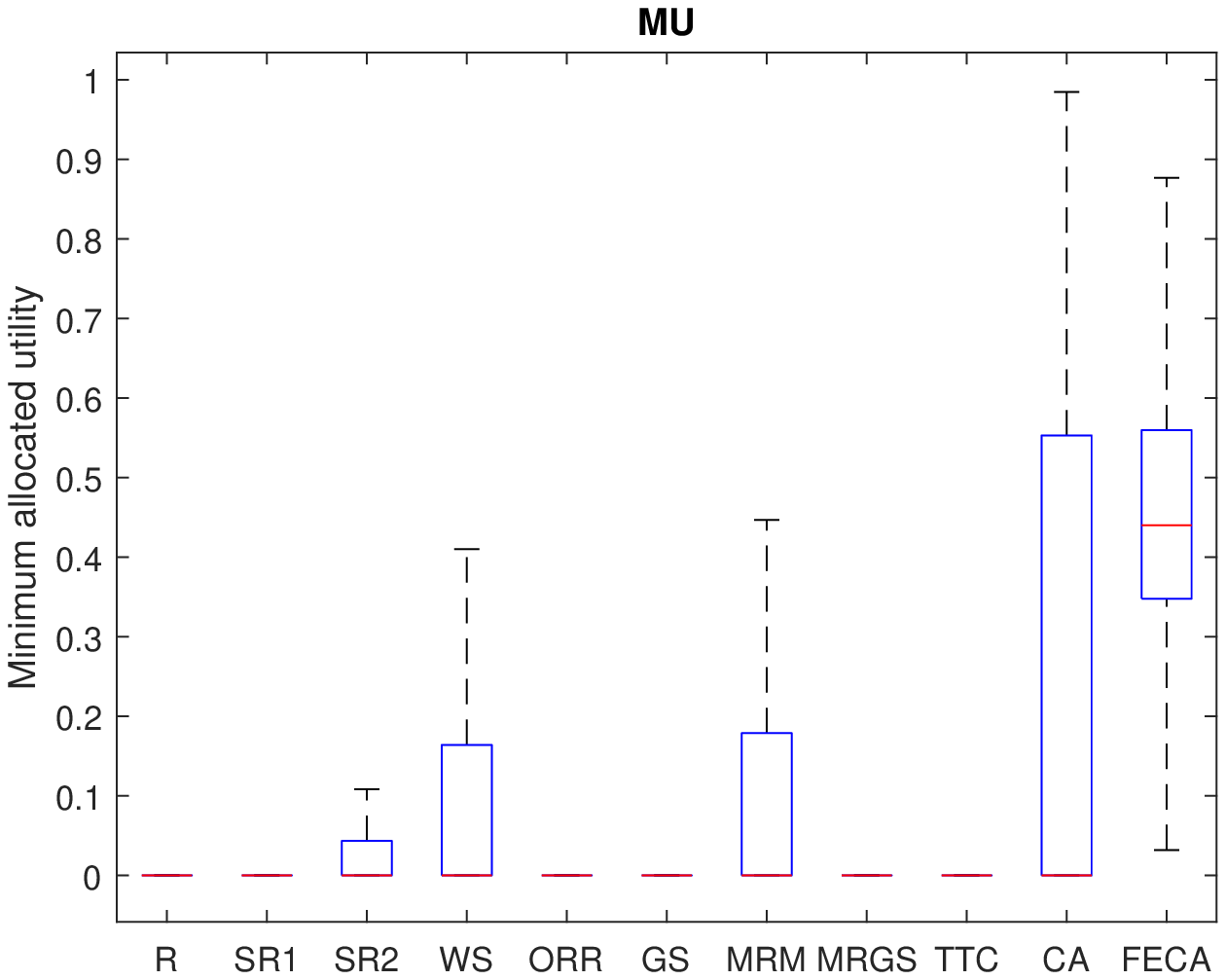}
         \caption{}
         \label{Fig_utility_MU_50red}
     \end{subfigure}     
     \hfill
     \caption{Minimal values of allocated capacity (a,c,e) and utility (b,d,f) context in cases I (a,b), II (c,d) and III (e,f). \label{Fig_MCMU}}
\end{figure}

However, the performance of these methods is remarkable from several aspects.
As shown in figure \ref{Fig_TCTU}, and in tables \ref{Tab_capacity_TC} and \ref{Tab_utility_TU}, the CA method, which simply aims for the highest \emph{overall} performance (in other words, not fairness-aware at any level),  
outperforms every other method, especially in the utility context, under all assumed levels of resource scarcity (i.e. in cases I, II and III).
If we look at figures \ref{Fig_F}, \ref{Fig_MCMU} and  table \ref{Tab_MCMU}, we may see that this overall overall performance implies a low resulting fairness in the capacity context. However, in the case of utility-based allocation, the $F_U$ values and resulting minimal utility levels of the CA are among the best.
Let us point out case III, where the resulting mean minimal utility value of the CA is higher by one order of magnitude compared to any other method (see table \ref{Tab_MCMU}), except the fairness-enhanced version of the algorithm (FECA), while it provides the best total utility in the same time.

The performance of the FECA allocation mechanism is even more remarkable. As tables \ref{Tab_capacity_F} and \ref{Tab_utility_F} show, independent of the context (capacity- or utility-based allocation), and resource scarcity, the mean
$F$ values are the highest in the case of this algorithm. Except for one case (case I of the capacity context), this is also true for the median values (as it may be seen in figure \ref{Fig_F} as well).

The convincing fairness properties of the FECA method are further supported by additional fairness-related measures.
If we look at the minimal values summarized in table \ref{Tab_MCMU}, we can see that regarding the mean minimal utility values ($MU$), the FECA outperforms every other method. Table \ref{Tab_nOC} furthermore shows, that in the case of 
utility-based allocation, it results in a very low number of outage events, even under resource scarcity (in case III it provides outstanding results).

But the really remarkable aspect of the results is that, as it is shown in
tables \ref{Tab_capacity_TC}, \ref{Tab_utility_TU} and figure \ref{Fig_TCTU}, the FECA algorithm is able to provide these very good fairness measures simultaneously with high system throughput/total utility. In contrast to selection based algorithms (including MRM) and MRGS, which are able to provide good fairness results if enough resources are available, but are not among the best performers when it comes to total system performance, the FECA gives a robust performance in both contexts.
 
 \subsection{Summary from a multi-objective viewpoint}
\begin{figure}[h!]
  \centering
  \includegraphics[width=.5\linewidth]{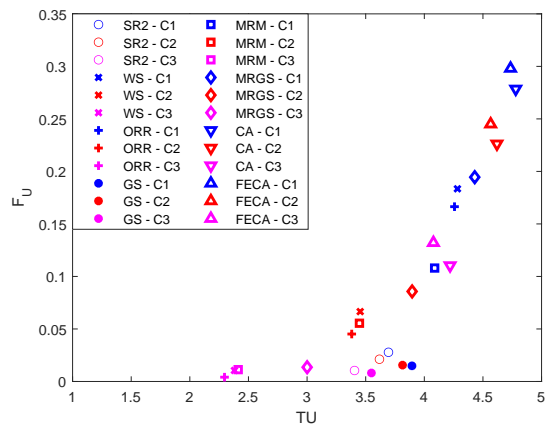}\\
  \caption{Total allocated utility versus fairness in the case of the various algorithms in the utility context (the low performance algorithms R, SR1, and TTC are not included for better visibility. C1, C2 and C3 refer to cases I, II and III respectively.)} \label{Fig_TU_vs_F}
\end{figure}

 We may summarize the most important points of the simulation results as follows (let us focus on the utility context, because taking into consideration minimal and maximal capacity values means a harder problem). Figure \ref{Fig_TU_vs_F} depicts the utility context simulation results in the space of total allocated utility and fairness. We can see that the CA-based algorithms perform well in this multi-objective context, thus they are able to provide high fairness values with high total throughput simultaneously, even when resource scarcity arises.
 
%\subsection{Additional aspects}
% can be added during the revsion!  
%One may distinguish between the discussed allocation methods according to multiple additional aspects as well. Regarding for example the implementation possibilities, we may note that CA and FECA definitely require a centralized implementation, while other methods, like GS have a potential to be implemented in a decentralized framework as well. 

%In addition, we must emphasize that both the CA and FECA methods are highly dependent on the appropriate preallocation algorithm. The many-to-many GS preallocation method used in this study has been shown to outperform more simple approaches like distance-based preallocation \cite{csercsik2021Heuristics}, but there is definitely a space for other, potentially more efficient preallocation methods for CA and FECA.

\section{Conclusions and future work}
\label{sec_conclusions}
% \subsection{Conclusions}

Overall, we have shown that in the case when the number of channels and tenants is not too high (around 15-20 and 6 respectively), but already makes the brute-force combinatorial optimization impossible, and one has to consider fairness and minimum/maximum user communication rate requirements as well, the proposed FECA algorithm shows a significant potential compared to other methods.
Regarding the total system throughput/total allocated utility, the CA also results in high values, but this simpler version is not able to ensure the appropriate fairness aspects.

%\subsection{Future work}

As it has been discussed in subsection \ref{subsec_FECA}, the basic optimization problem of FECA may be infeasible. In this study we used the approach of
iteratively relaxing all constraints describing the minimal resulting values.
Instead of this simple method, more sophisticated algorithms may increase the performance of FECA even further. One may look e.g. each time for the most critical constraint, and relax the requirements only for the tenant involved in the respective constraint. %while keeping the minimal capacity/utility levels of other tenants unaffected.

% future work ideas (currently I see that there's no space for these):

% strong interference - externalities model: If an assignment explicitly influences the resulting capacity/utility values of other assignments

%  smart relaxation of the FECA: If the FECA is infeasible, relax the fairness constraints in a smart way

% convex combinatorial auction in the context of time-sharing (?)

% find appropriate wireless applications for the TTC

% Both elastic and non-elastic traffic. 

%\section{Acknowledgments}
%This work has been supported by the Hungarian Academy of Sciences under its Momentum Programme LP2021-2 and by the Fund and by K 131 545 
%of the Hungarian National Research, Development and Innovation Office. 
%This work has been supported by the Fund and by K 131 545 of the Hungarian National Research, Development and Innovation Office.
%D\'{a}vid Csercsik is a grantee of the Bolyai scholarship program of the Hungarian Academy of Sciences.

%The work of E. Jorswieck is partly funded by the German Research Foundation (DFG) under grant JO 801/24-1 and by the Federal Ministry of Education and Research (BMBF) within the 6G Research and Innovation Center (6G-RIC) under support code 16KISK031.  

\bibliographystyle{IEEEtran}
\bibliography{GS_RR}

%\section*{Appendix A: Simulation parameters}

%Following \cite{Hossler2019matching}, the transmission parameters of the used URLLC model were as described in table \ref{Tab_params}.

  %\begin{table}[h!]
  %\begin{center}
  %\begin{tabular}{|c|c|c|c|c|c|}
    %\hline
%    % after \\: \hline or \cline{col1-col2} %\cline{col3-col4} ...
    %channel bandwidth & $B$ & 20 MHz & Rician Factor %reference value& $K_{ref}$ & 14.1 dB \\
    %reference distance & $d_0$ & 15m & Interference %power & $P_I$ & -50 dBm\\
    %reference path loss & $PL(d_0)$ & 70.28 dB & %outage probability threshold & $\varepsilon$ & %$10^{-9}$ \\
    %PL exponent & $\delta$ & 2 & & & \\
    %\hline
%  \end{tabular}
  %\end{center}
  %\caption{Simulation parameters \label{Tab_params}}
  %\end{table}

% ===== from here, content will be included only in the ArXive version
\newpage
\section*{Appendix A: Utility function example}

\begin{figure}[h!]
  \centering
  \includegraphics[width=.7\linewidth]{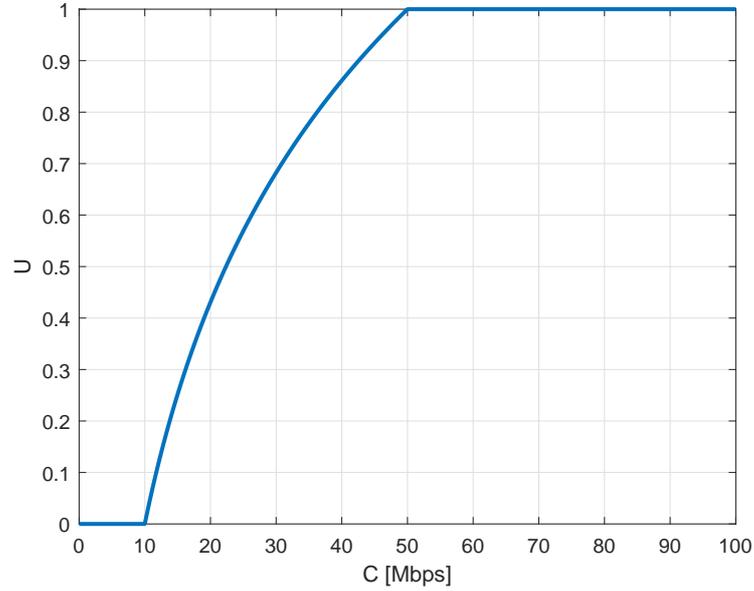}\\
  \caption{Example of the proposed utility function in (\ref{utility_fncn}), assuming
$C_k^{min}=10$ Mbps and $C_k^{max}=50$ Mbps.} \label{utility_fncn_example}
\end{figure}

\section*{Appendix B: Bid matrix example}

Let us consider the bid matrix described in Eq. \ref{bid_matrix_example}.

\begin{equation}
\left(
  \begin{array}{cccccc}
    1 & 1 & 0 & 0 & 15   & 1 \\
    1 & 0 & 0 & 1 & 13   & 1 \\
    0 & 1 & 0 & 1 & 14   & 1 \\
    1 & 1 & 0 & 1 & 42   & 1 \\
    1 & 1 & 0 & 0 & 9    & 2 \\
    1 & 0 & 1 & 0 & 17   & 2 \\
    0 & 1 & 1 & 0 & 18   & 2 \\
    1 & 1 & 1 & 0 & 31   & 2 \\
  \end{array}
\right)
\label{bid_matrix_example}
\end{equation}

In this case we have altogether 8 bids, submitted by two players for subsets of 4 available channels ($CH^1=\{1,2,4\},~~CH^2=\{1,2,3\}$).
In this case we omitted single-channel bids.
Even in the case of this simple example, if both tenants would place bids for all possible subsets, the matrix would have $2\cdot(2^4-1)$ rows.
In general, as in the process of the combinatorial auction a binary acceptance variable is assigned to each bid (row), the method would boil down in this case practically to brute-force optimization (it is sure that we could find the optimal assignment), and would require enormous computing power.

If we consider the bid matrix described in Eq. \ref{bid_matrix_example} , the constraints of the CA problem may be easily formalized as inequality constraints.
In the case of the example bid matrix described in eq. (\ref{bid_matrix_example}), the first set of constraints is summarized in eq. (\ref{constr_example_1}), as

\begin{align}
x_1+x_2+x_3+x_4 &\leq 1 ~~~ x_5+x_6+x_7+x_8 \leq 1~~,
\label{constr_example_1}
\end{align}
where $x_i$ is the (binary) acceptance indicator of bid $i$.

The second set of constraints (see eq. (\ref{constr_example_2})), may be derived from the respective columns of the bid matrix, as
\begin{align}
x_1+x_2+x_4+x_5+x_6+x_8 & \leq  1 ~~~ x_1+x_3+x_4+x_5+x_7+x_8  \leq  1 \nonumber \\
x_6+x_7+x_8 & \leq  1~~~~ x_2+x_3+x_4  \leq  1 \label{constr_example_2}
\end{align}

% todo: suitable journal or special issue? EJ
% IEEE Trans. or Journal? 
 
 %==========================
 
\section*{Appendix C: Additional details of simulation results}

\begin{table}[h!]
    \centering
    \begin{tabular}{|c|c|c|c|c|c|c|c|c|c|c|c|c|}
    \hline
    & & R & SR1 & SR2 & WS & ORR & GS & MRM & MRGS & TTC & CA & FECA\\ \hline
\multirow{ 2}{*}{Case I} & median($TC$) & 53.09    &     70.21  &       70.00  &       60.36    &     63.60    &     99.26      &   72.54    &     73.17  &  67.85    &   \textbf{102.97}    &     90.53 \\ 
& mean($TC$) & 53.88    &     72.59     &    72.62      &   61.84  &       65.28 &        101.96    &     74.65     &    74.72 &   70.19    &    \textbf{105.95}     &    93.36 \\ \hline
\multirow{ 2}{*}{Case II} & median($TC$) &  36.05     &    47.76      &   67.35 &         43.98    &     48.07       &  90.13     &    54.97     &    61.08    & 58.08    &     \textbf{93.59}      &   81.85 \\ 
& mean($TC$) & 36.73   &      50.88    &     69.75     &    45.64    &     49.72 &         92.67    &     57.90     &    62.74   & 60.02     &    \textbf{96.66}      &   84.92 \\ \hline
\multirow{ 2}{*}{Case III} & median($TC$) &  20.26   &      27.11   &      60.35&         26.43      &   30.82      &   75.66   &      32.68   &      43.81 & 42.42    &     \textbf{79.92}   &       66.12 \\ 
& mean($TC$) & 20.92   &      30.25     &    62.97    &     28.15     &    32.89&         78.52     &    37.56     &    45.71 & 44.66    &     \textbf{82.79}   &      68.91 \\ \hline
    \end{tabular}
        \hspace{1.ex}
    \caption{Median and mean values of total allocated capacity ($TC$) in cases I, II and III (capacity-based allocation)}
    \label{Tab_capacity_TC}
\end{table}

\begin{table}[h!]
    \centering
    \begin{tabular}{|c|c|c|c|c|c|c|c|c|c|c|c|c|}
    \hline
    & & R & SR1 & SR2 & WS & ORR & GS & MRM & MRGS & TTC & CA & FECA\\ \hline
\multirow{ 2}{*}{Case I} & median($TU$) &  3.401 &    3.715 &    3.706 &    4.350 &    4.329 &    3.892 &    4.152 &    4.522 &    3.808 &    \textbf{4.814} &    4.810 \\ 
& mean($TU$) & 3.389 &    3.693 &    3.694 &    4.282 &    4.258 &    3.895 &    4.088 &    4.430 &    3.743 &    \textbf{4.780} &    4.736 \\ \hline
\multirow{ 2}{*}{Case II} & median($TU$) &  2.441 &    2.750 &    3.643 &    3.522 &    3.432 &    3.813 &    3.537 &    3.960 &    2.500 &    \textbf{4.647} &    4.636 \\ 
& mean($TU$) & 2.445 &    2.751 &    3.617 &    3.453 &    3.381 &    3.814 &    3.447 &    3.896 &    2.499 &    \textbf{4.619} &    4.566 \\ \hline
\multirow{ 2}{*}{Case III} & median($TU$) &  1.423 &    1.683 &    3.423 &    2.312 &    2.293 &    3.579 &    2.336 &    3.023 &    1.290 &    \textbf{4.239 }&    4.132 \\ 
& mean($TU$) & 1.439 &   1.705 &    3.405 &    2.380 &    2.296 &    3.549 &    2.412 &    3.001 &    1.331 &    \textbf{4.220 }&    4.078 \\ \hline
    \end{tabular}
        \hspace{1.ex}
    \caption{Median and mean values of total allocated utility ($TU$) in cases I, II and III (utility-based allocation)}
    \label{Tab_utility_TU}
\end{table}

%==============================================================================
% Fairness tables

\begin{table}[h!]
    \centering
    \begin{tabular}{|c|c|c|c|c|c|c|c|c|c|c|c|c|}
    \hline
    & & R & SR1 & SR2 & WS & ORR & GS & MRM & MRGS & TTC & CA & FECA\\ \hline
\multirow{ 2}{*}{Case I} & median($F_C$) & 0.019 &    0.050 &    0.052 &    0.863 &    0.838 &         0   & 0.560 &    1.543 &    1.012 &         0   & \textbf{1.329} \\ 
& mean($F_C$) & 0.218 &    0.865 &    0.845 &    3.233 &    3.288 &    1.985 &    2.886 &    5.621 &    4.881 &    0.698 &    \textbf{7.916} \\ \hline
\multirow{ 2}{*}{Case II} & median($F_C$) & 0.000 &    0.000 &    0.029 &    0.103 &    0.052 &         0   & 0.131 &    0.275 &    0.219 &         0   & \textbf{0.846}\\ 
& mean($F_C$) & 0.014 &   0.070 &    0.586 &    0.687 &    0.597 &    1.270 &    1.133 &    1.667 &    1.778 &    0.643 &    \textbf{4.916} \\ \hline
\multirow{ 2}{*}{Case III} & median($F_C$) &  0.000 &    0.000 &    0.002 &    0.001 &    0.001 &         0   & 0.002 &    0.007 &    0.005 &         0   & \textbf{0.255} \\ 
& mean($F_C$) & 0.000 &    0.002 &    0.243 &   0.062 &    0.048 &    0.402 &    0.192 &    0.177 &    0.255 &    0.168 &   \textbf{ 1.632} \\ \hline
    \end{tabular}
        \hspace{1.ex}
    \caption{Median and mean values of $F_C$ in cases I, II and III (capacity-based allocation, values are normalized by $10^6$)}
    \label{Tab_capacity_F}
\end{table}

\begin{table}[h!]
    \centering
    \begin{tabular}{|c|c|c|c|c|c|c|c|c|c|c|c|c|}
    \hline
    & & R & SR1 & SR2 & WS & ORR & GS & MRM & MRGS & TTC & CA & FECA\\ \hline
\multirow{ 2}{*}{Case I} & median($F_U$) & 0.004 &   0.007 & 0.006 &    0.136 &    0.1189 &     0  &  0.082 &    0.149 &    0.046 &    0.239 &    \textbf{0.2394} \\ 
& mean($F_U$) &  0.018 &    0.028 &    0.028 &    0.183 &    0.166 &    0.015 &    0.108 &    0.195 &    0.081 &    0.279 &    \textbf{0.298} \\ \hline
\multirow{ 2}{*}{Case II} & median($F_U$) &       0  &       0  &  0.004 &    0.032 &         0     &    0  &  0.031 &    0.039 &          0 &    0.188 &   \textbf{ 0.190} \\ 
& mean($F_U$) & 0.002 &    0.003 &    0.021 &    0.067 &    0.045 &    0.016 &    0.056 &    0.086 &    0.007 &    0.226 &    \textbf{0.245} \\ \hline
\multirow{ 2}{*}{Case III} & median($F_U$) &      0     &    0 &        0     &    0    &     0 &        0      &   0  &       0    &     0     &    0 &   \textbf{0.084} \\ 
& mean($F_U$) & 0.000 &    0.000 &    0.010 &    0.011 &    0.004 &    0.008 &    0.011 &    0.014 &    0.000 &    0.110 &    \textbf{0.132} \\ \hline
    \end{tabular}
        \hspace{1.ex}
    \caption{Median and mean values of $F_U$ in cases I, II and III (utility-based allocation)}
    \label{Tab_utility_F}
\end{table}
%===============================================================================
% Minimal values

\begin{figure}[h!]
  \centering
  \includegraphics[width=.47\linewidth]{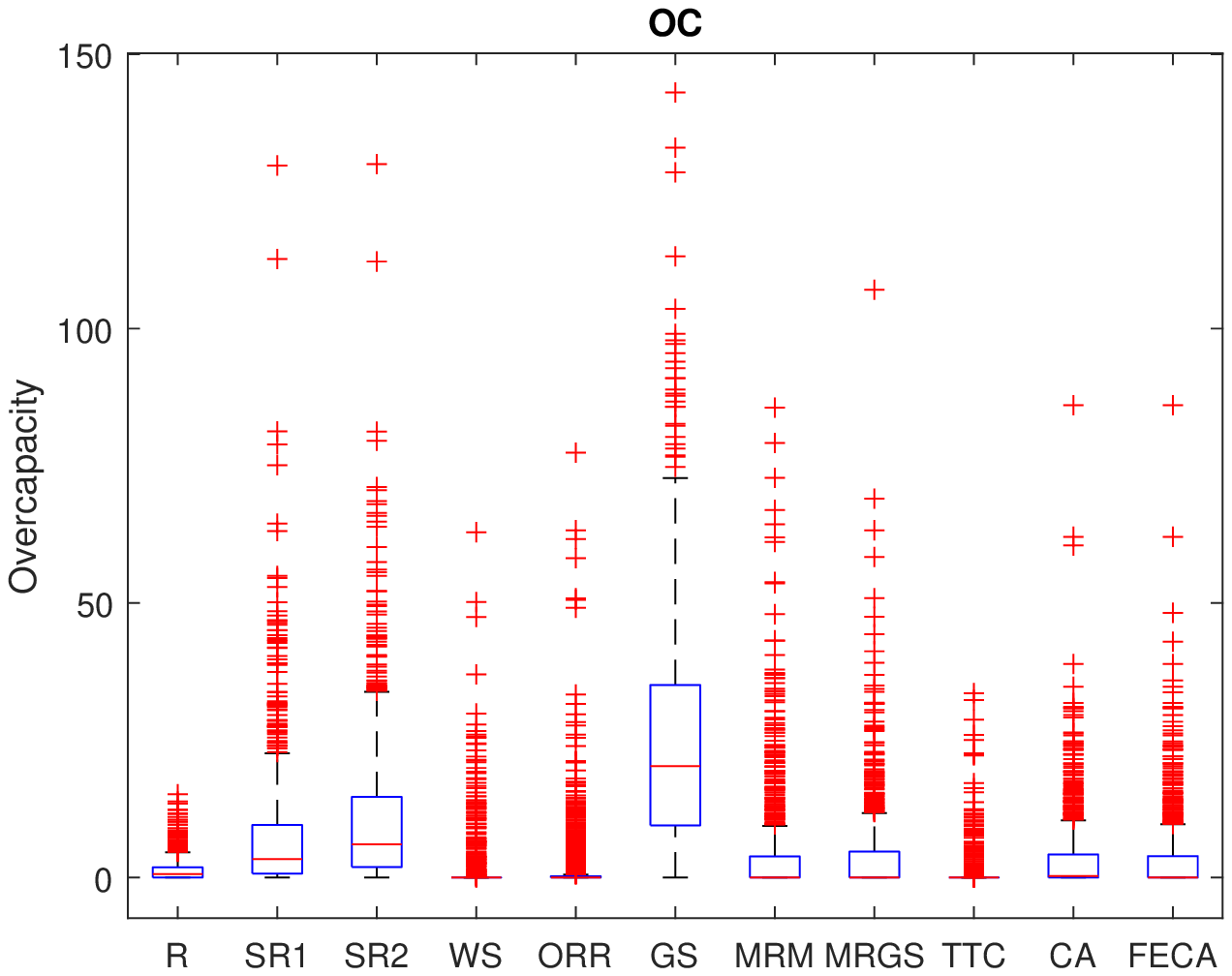}
  \includegraphics[width=.47\linewidth]{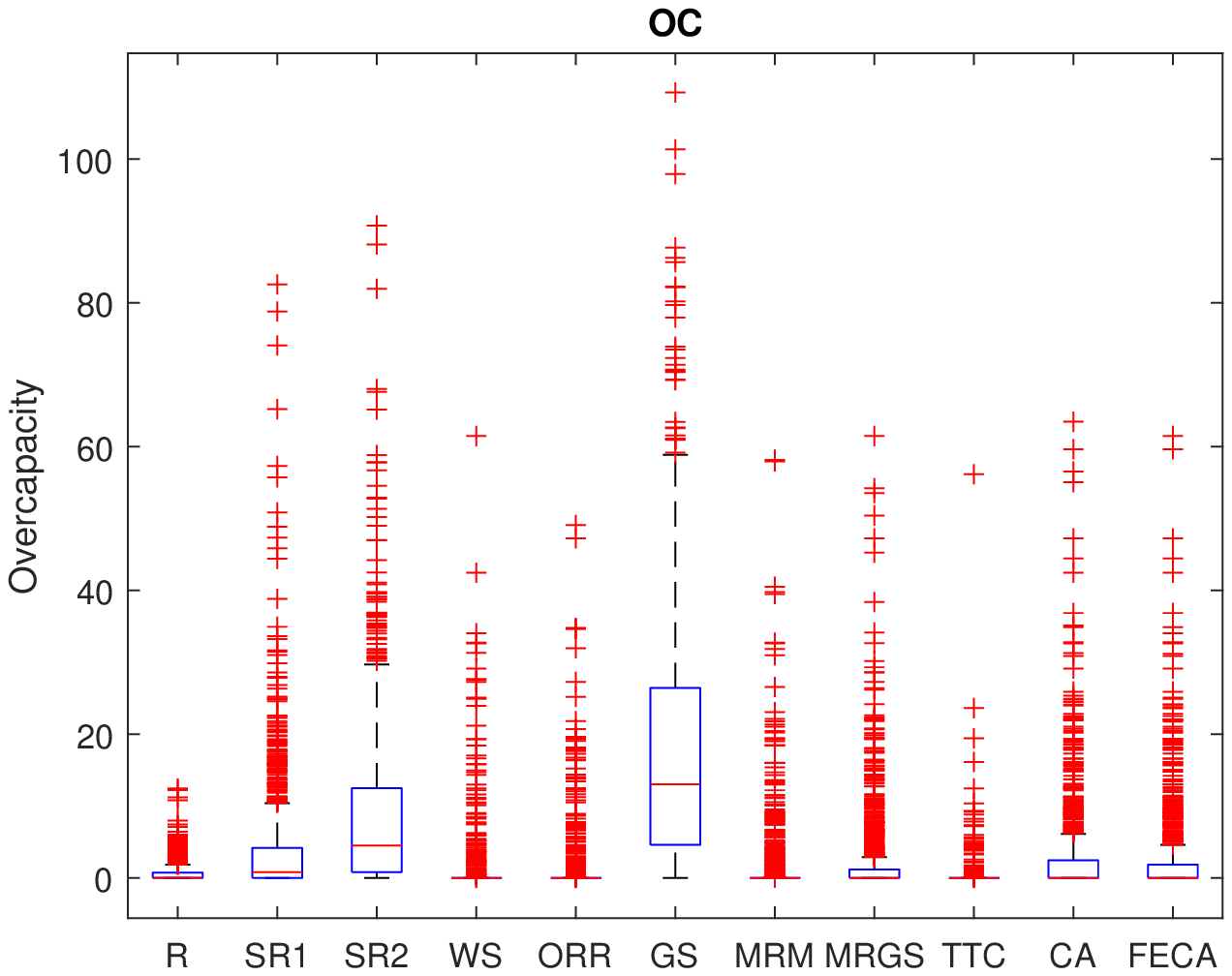}
  \caption{Overcapacity [Mbps] in Cases II and III of the utility context} \label{Fig_utility_OC_2550red}
\end{figure}

%===============================================================================
% comp time

\begin{figure}
     \centering
     \begin{subfigure}[b]{0.48\textwidth}
         \centering
         \includegraphics[width=\textwidth]{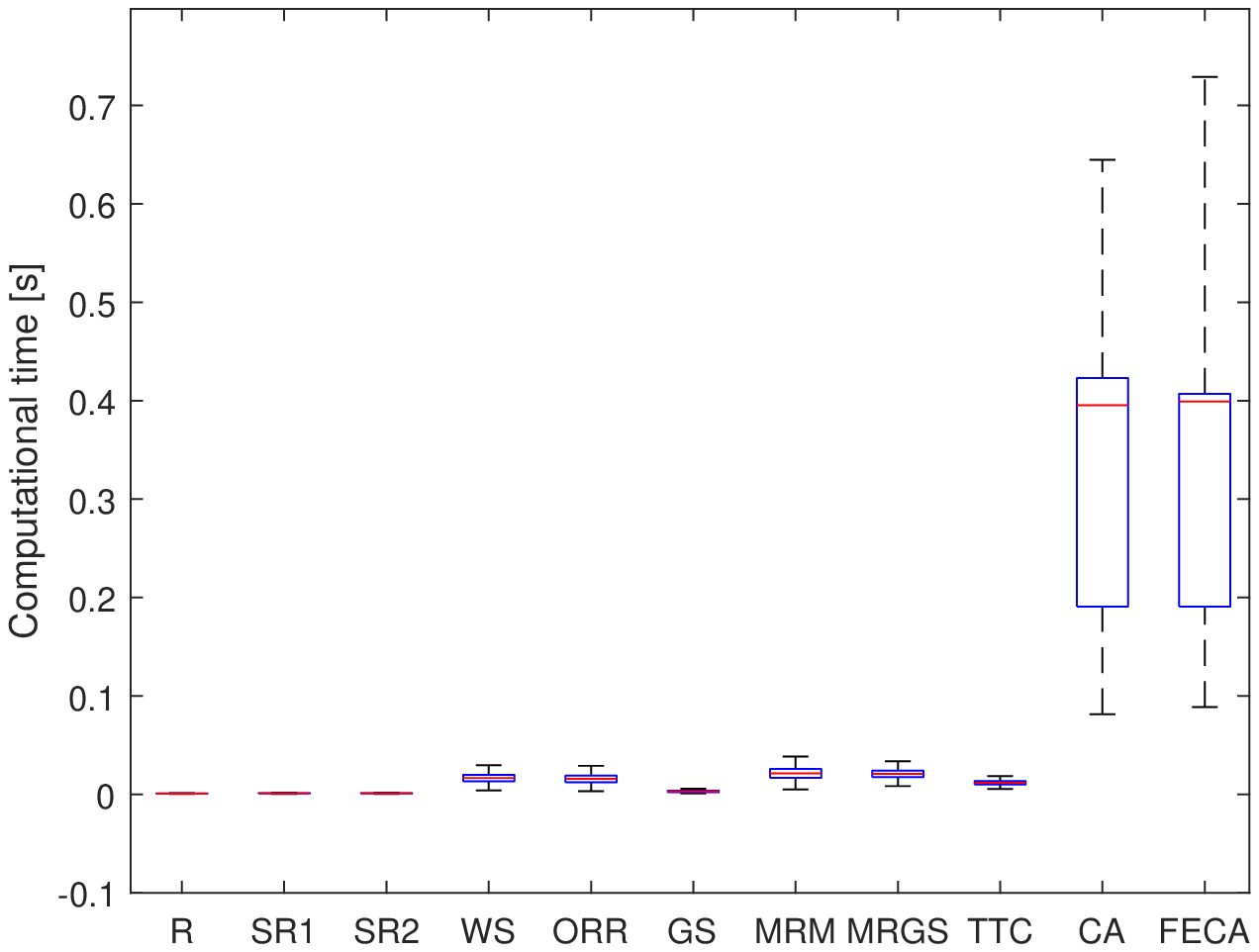}
         \caption{}
         \label{Fig_capacity_t_0red}
     \end{subfigure}
     \hfill
     \begin{subfigure}[b]{0.48\textwidth}
         \centering
         \includegraphics[width=\textwidth]{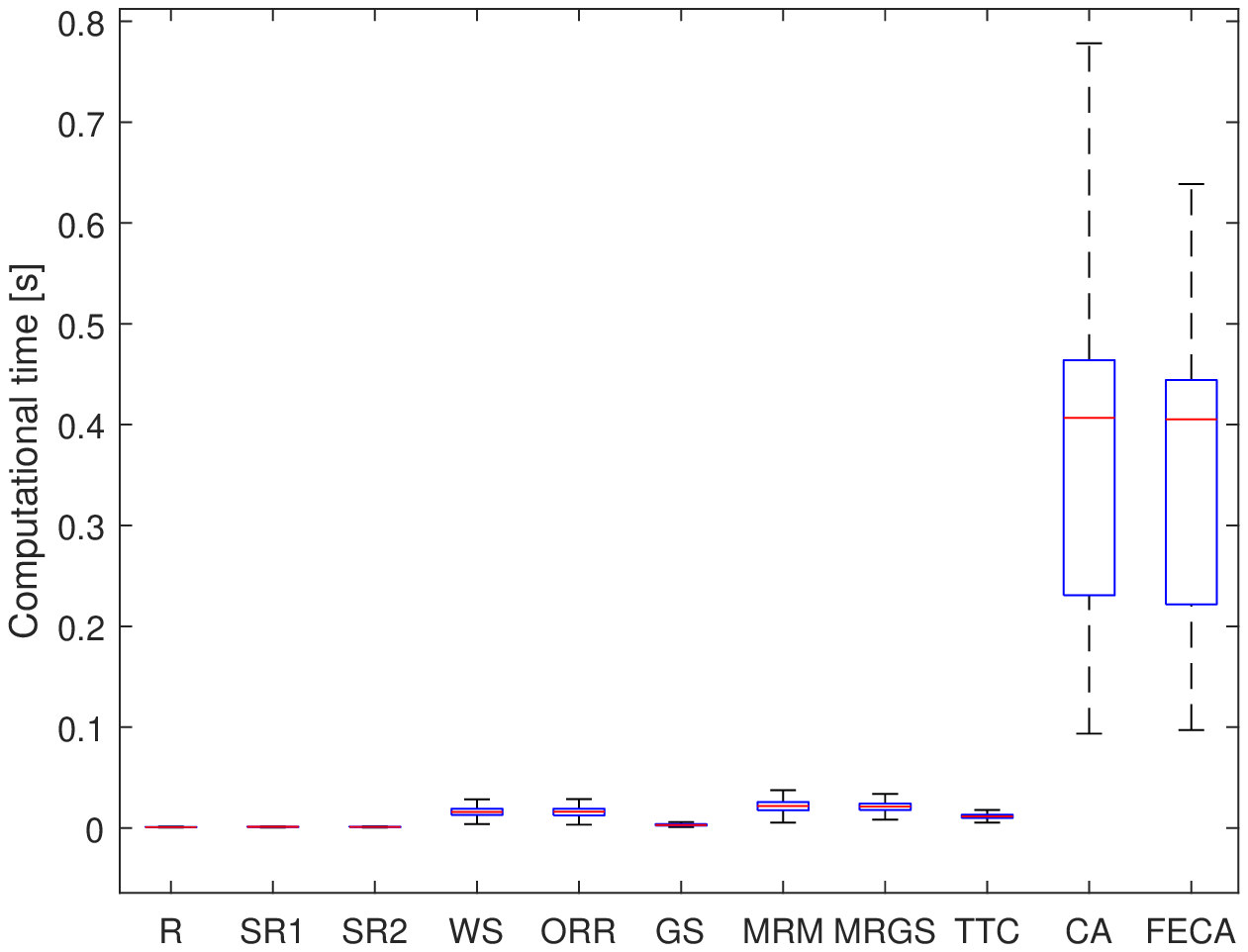}
         \caption{}
         \label{Fig_utility_t_0red}
     \end{subfigure}
     \hfill
     \begin{subfigure}[b]{0.48\textwidth}
         \centering
         \includegraphics[width=\textwidth]{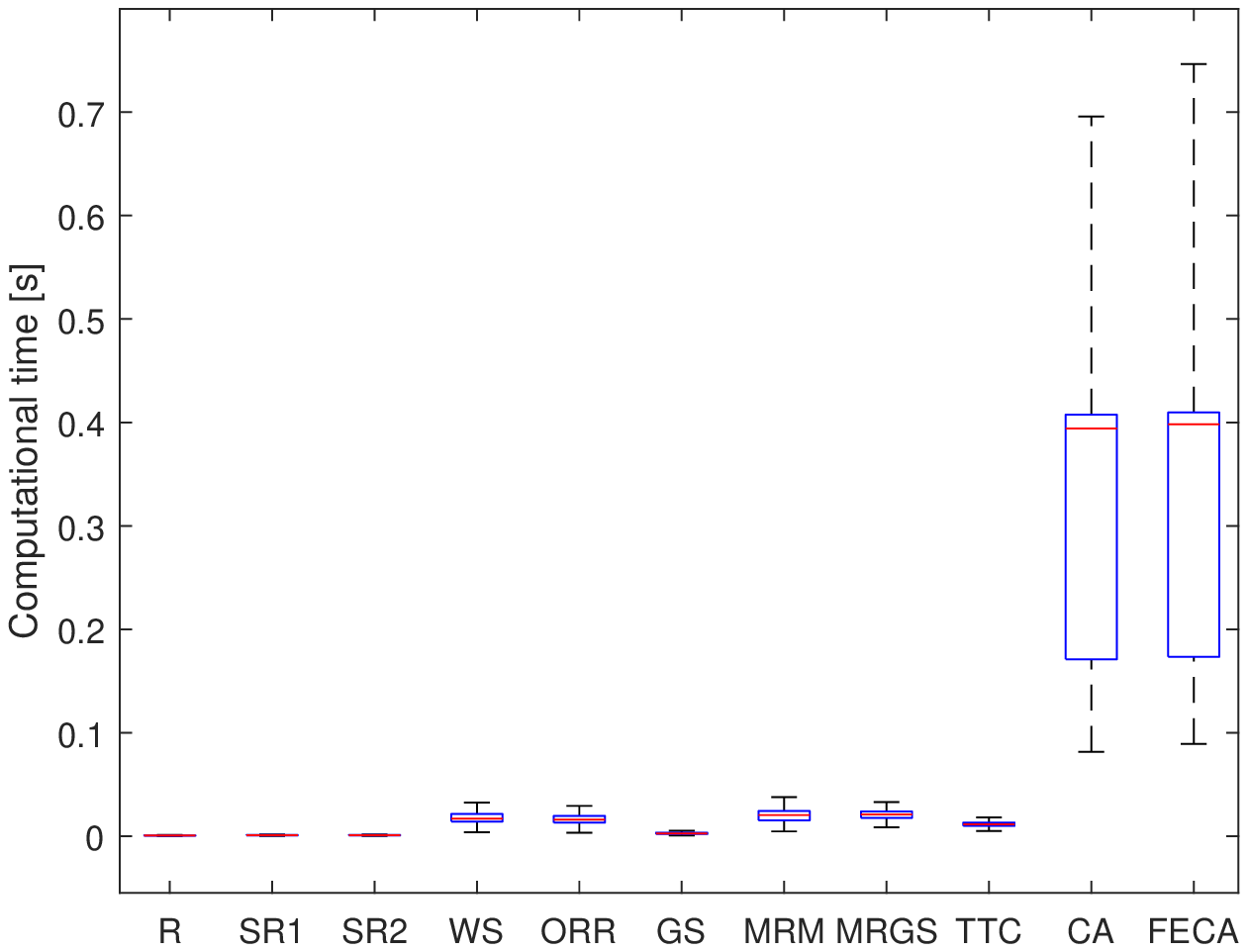}
         \caption{}
         \label{Fig_capacity_t_25red}
     \end{subfigure}
     \hfill
     \begin{subfigure}[b]{0.48\textwidth}
         \centering
         \includegraphics[width=\textwidth]{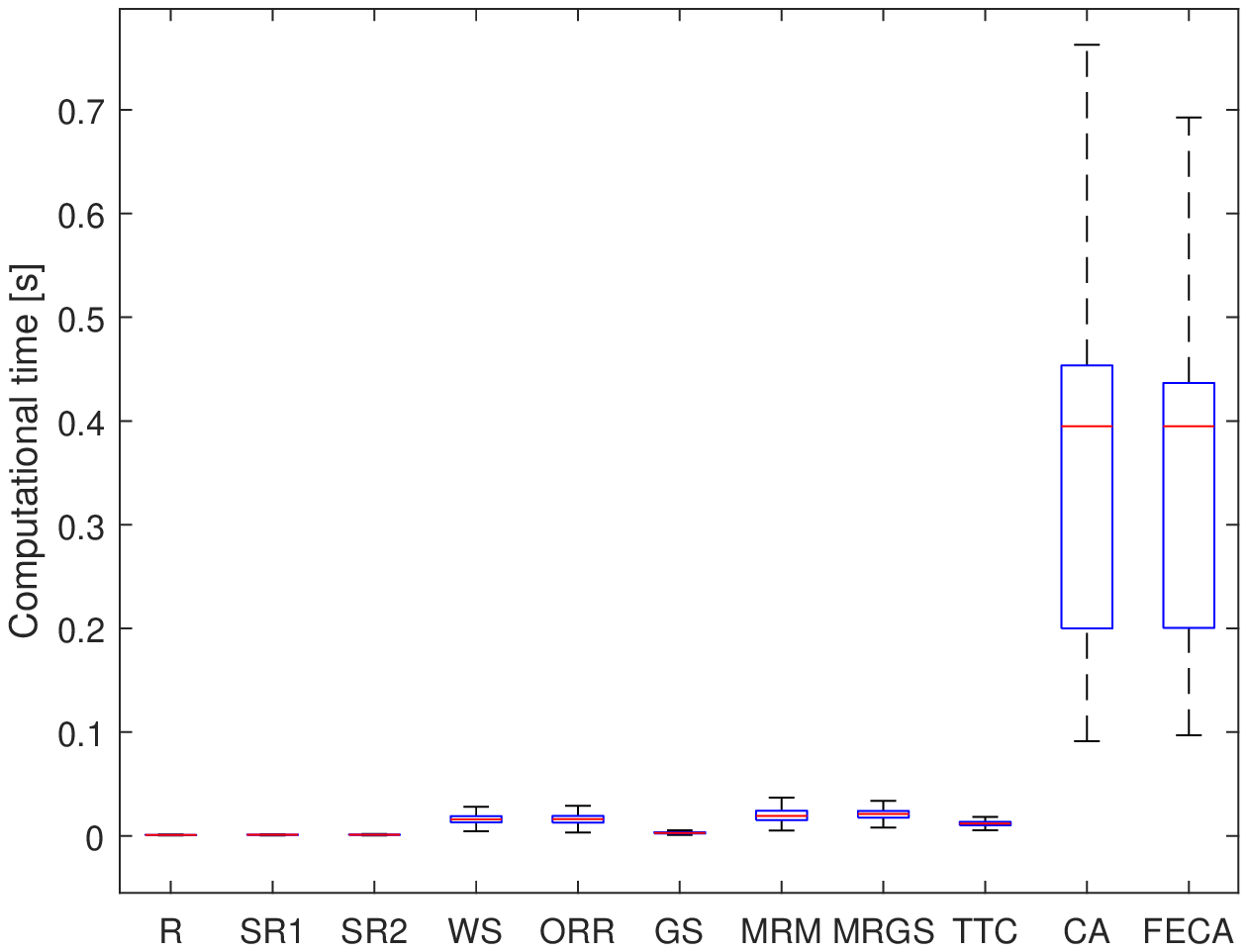}
         \caption{}
         \label{Fig_utility_t_25red}
     \end{subfigure}
     \hfill     
     \begin{subfigure}[b]{0.48\textwidth}
         \centering
         \includegraphics[width=\textwidth]{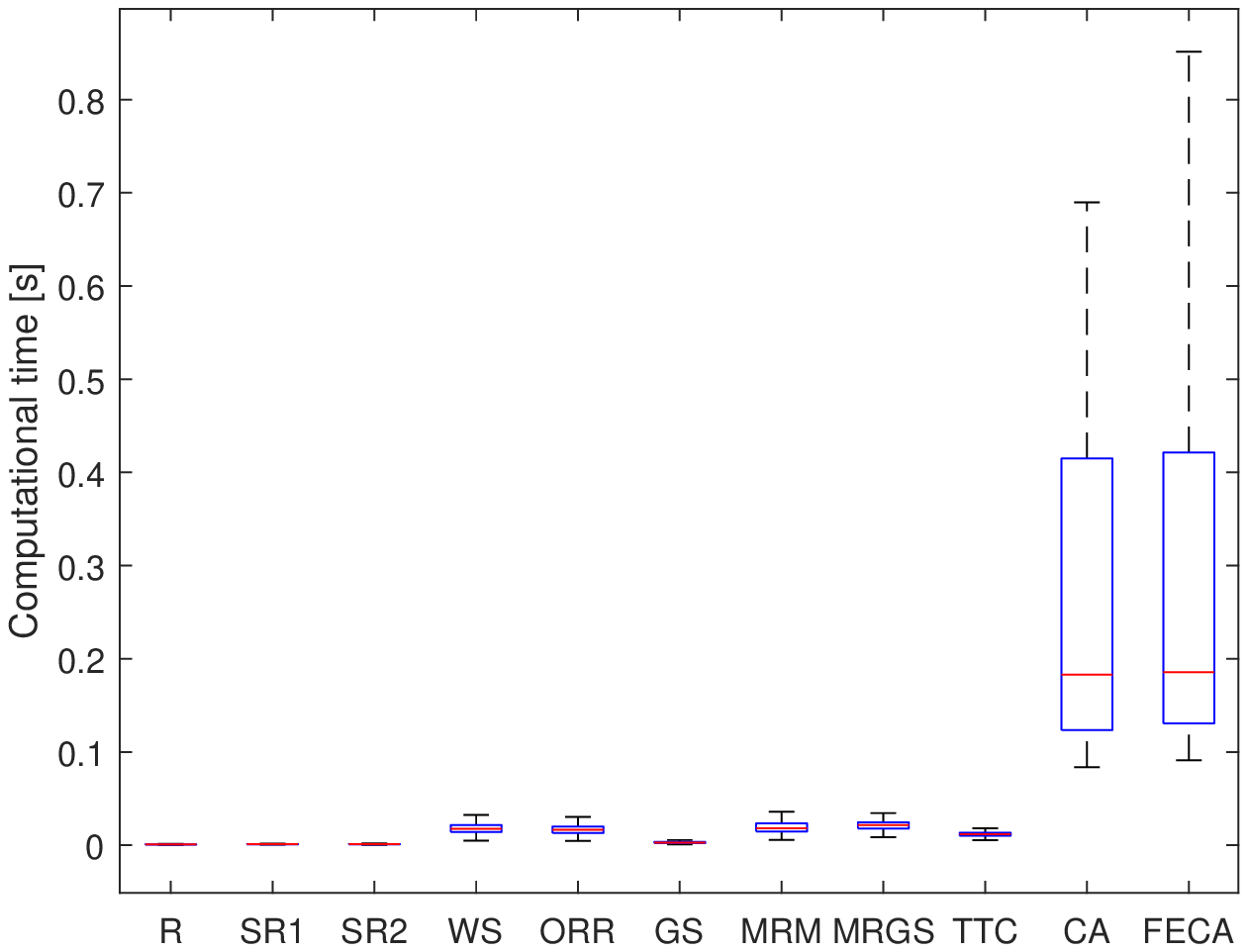}
         \caption{}
         \label{Fig_capacity_t_50red}
     \end{subfigure}
     \hfill
     \begin{subfigure}[b]{0.48\textwidth}
         \centering
         \includegraphics[width=\textwidth]{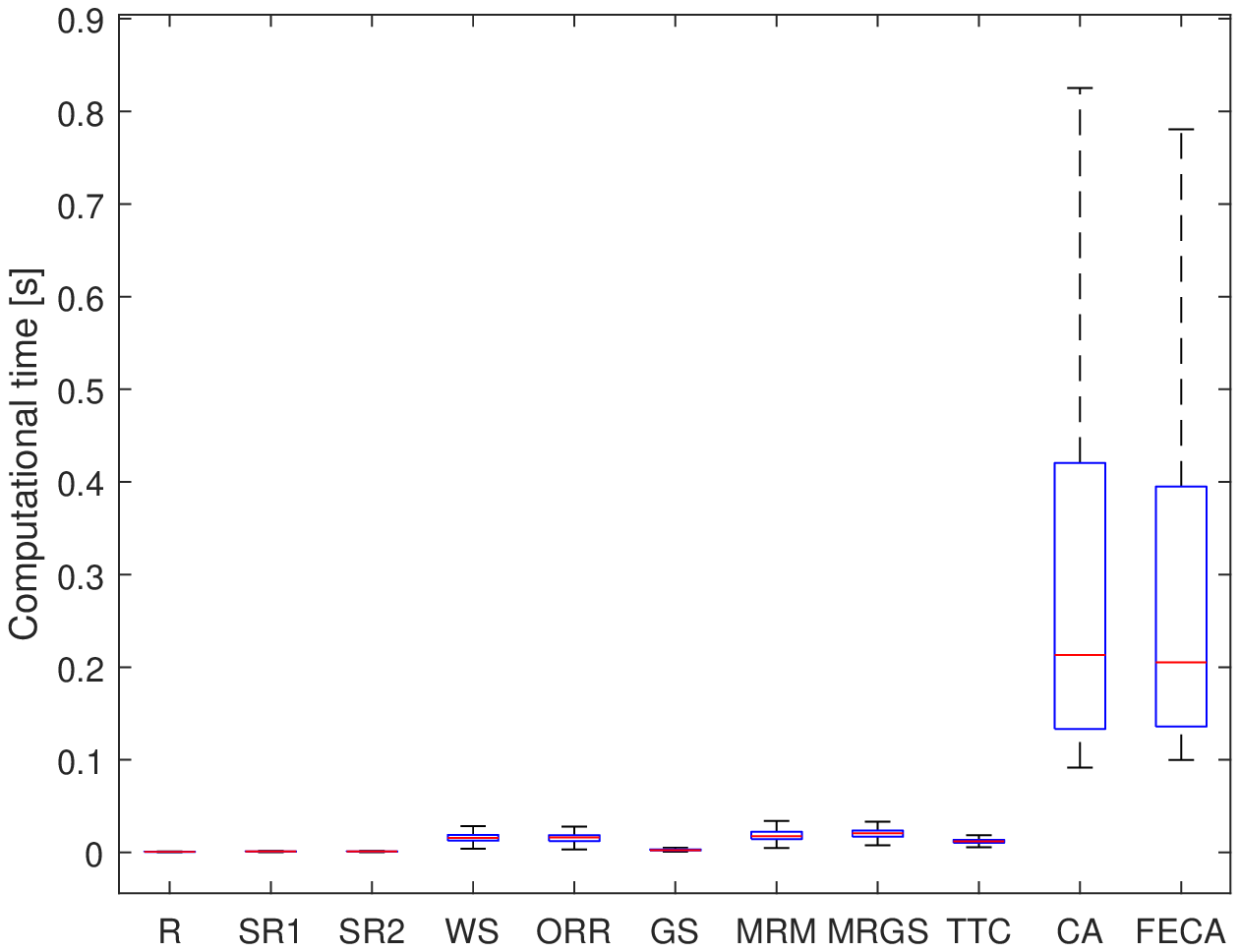}
         \caption{}
         \label{Fig_utility_t_50red}
     \end{subfigure}     
     \hfill
     \caption{Computational time in cases I (a,b), II (c,d) and III (e,f) in the case of capacity-based (a,c,e) and utility-based (b,d,f) allocation. \label{Fig_t}}
\end{figure}

\end{document}